\documentclass[amssymb,prb,twocolumn,superscriptaddress,floats,showkeys,amsmath]{revtex4-2}

\usepackage{siunitx}
\usepackage[version=3]{mhchem} 
\usepackage{bm}
\usepackage[utf8]{inputenc}
\usepackage[T1]{fontenc}
\usepackage{graphicx}
\usepackage{upgreek}
\usepackage{color}
\usepackage{ulem}
\usepackage{hyperref} 
\hypersetup{colorlinks, citecolor=blue, filecolor=blue, linkcolor=blue, urlcolor=blue}
\usepackage{sidecap}
\usepackage{amssymb}
\usepackage[caption=false]{subfig}
\usepackage{xr}
\usepackage{xcolor}
\definecolor{darkgreen}{RGB}{0,80,0}

\def \FUW{Faculty of Physics, University of Warsaw, 02-093 Warsaw, Poland}

\def \Watanabe{Research Center for Electronic and Optical Materials, National Institute for Materials Science, 1-1 Namiki, Tsukuba 305-0044, Japan}
\def \Taniguchi{Research Center for Materials Nanoarchitectonics, National Institute for Materials Science,  1-1 Namiki, Tsukuba 305-0044, Japan}
\def \Sapienza{Physics Department, Sapienza University of Rome, 00185 Rome, Italy}

\def \LNCMI{Laboratoire National des Champs Magnétiques Intenses, CNRS-UGA-UPS-INSA-EMFL, 38042 Grenoble, France}

\def \Centera{CENTERA, CEZAMAT, Warsaw University of Technology, 02-822 Warsaw, Poland}
\def \Charles{Department of Condensed Matter Physics, Faculty of Mathematics and Physics, Charles University, Prague, Czech Republic}
\def\afiliacjamehdiego {Université Paris-Saclay, CEA, CNRS, SPEC, 91191 Gif-sur-Yvette, France}
\def\afiliacjamehdi {Université Paris-Saclay, ONERA, CNRS, Laboratoire d'étude des microstructures (LEM), 92322 Châtillon, France}

\def \Dipankar{Institute for Functional Intelligent Materials, National University of Singapore, 117544, Singapore}

\begin{document}

\title{Magnetic-Field-Induced Darkening of Bright Excitons \\in a MoWSe$_2$ Monolayer Alloy}

\author{Katarzyna Olkowska-Pucko}
\email{katarzyna.olkowska-pucko@fuw.edu.pl}
\affiliation{\FUW}
\author{Artur O. Slobodeniuk}
\email{artur.slobodeniuk@matfyz.cuni.cz}
\affiliation{\Charles}
\affiliation{\Centera}
\author{Mehdi Arfaoui}
\affiliation{\afiliacjamehdiego}
\affiliation{\afiliacjamehdi}
\author{Elena Blundo}
\affiliation{\Sapienza}
\author{Natalia Zawadzka}
\affiliation{\FUW}
\author{Łucja~Kipczak}
\affiliation{\FUW}
\author{Jan Szpakowski}
\affiliation{\FUW}
\author{Diana~Vaclavkova}
\affiliation{\LNCMI}
\author{Dipankar~Jana}
\affiliation{\LNCMI}
\affiliation{\Dipankar}
\author{Piotr Kapu\'sci\'nski}
\affiliation{\FUW}
\affiliation{\LNCMI}
\author{Kenji~Watanabe}
\affiliation{\Watanabe}
\author{Takashi Taniguchi}
\affiliation{\Taniguchi}
\author{Clement~Faugeras}
\affiliation{\LNCMI}
\author{Marek Potemski}
\affiliation{\FUW}
\affiliation{\LNCMI}
\affiliation{\Centera}
\author{Adam Babi\'nski}
\affiliation{\FUW}
\author{Antonio Polimeni}
\affiliation{\Sapienza}
\author{Maciej R. Molas}
\email{maciej.molas@fuw.edu.pl}
\affiliation{\FUW}

\begin{abstract}
Magnetic fields are widely used to brighten spin-forbidden dark excitons, but the inverse effect on an optically bright exciton has remained largely unexplored. 
We demonstrate the complementary phenomenon of magnetic-field-induced darkening of a bright exciton in Mo$_{0.68}$W$_{0.32}$Se$_2$.
At 4.2 K, an in-plane magnetic field up to 25 T activates a dark-exciton line 14.2$\pm$1.3 meV below the bright exciton while suppressing the bright-exciton emission by about 85\%.
A bright-dark mixing model quantitatively captures the coupled energy and intensity evolution with an effective in-plane $g$ factor of $|g_{\parallel}|=5.2\pm0.2$. 
Alloying is crucial to the effect, as it brings the system close to the bright-dark crossover, where a small bright-dark splitting is combined with unusually strong in-plane magnetic coupling.
This places the system in a regime where field-induced level repulsion raises the bright-like branch and strongly depletes its thermal population.
These results identify a general route to magnetic control of exciton populations through engineering of the bright-dark hierarchy.

\end{abstract}

\maketitle

\textit{Introduction--} The optical activation (brightening) of dark excitons, electron-hole states whose direct radiative recombination is suppressed by spin-selection rules, is a well-established phenomenon in semiconductor nanostructures.
Dark excitonic states and their coupling to optically active counterparts have been extensively investigated in conventional quantum wells~\cite{Kurdyubov2021} and semiconductor quantum dots, where dark-exciton brightening can be induced by magnetic fields, structural symmetry breaking, or coupling to magnetic impurities~\cite{Bayer2000, Goryca2010, Huo2017, Germanis2018}.
In such systems, dark excitons provide access to otherwise hidden excitonic fine structure and play an important role in spin relaxation, exchange interactions, and the formation of long-lived excitonic reservoirs.
More recently, dark-exciton physics has gained increasing attention in layered semiconductors, including layered metal-halide perovskites~\cite{Dyksik2021,Posmyk2022,Thompson2024} and atomically thin van der Waals (vdW) materials, such as InSe~\cite{Chen2024InSe}, magnetic CrSBr~\cite{Sears2025CrSBr}, and, most prominently, semiconducting transition-metal dichalcogenides (S-TMDs)~\cite{Echeverry2016, Slobodeniuk2016, Molas_2017_DX_BDX, Zhang2017, Wang2018, Molas2019,Lu_2020, Robert2020,Pucko_2023,Cianci2023}.

Monolayers (MLs) of Mo- and W-based S-TMDs provide a particularly rich platform for exploring bright- and dark-exciton physics.
Owing to the absence of inversion symmetry and strong spin-orbit coupling, the electronic bands at the inequivalent K$^{\pm}$ valleys are spin split. 
Consequently, the sign of the conduction-band (CB) spin splitting determines whether the lowest intravalley excitonic transition is spin allowed or spin forbidden~\cite{Liu2013,Kosmider2013,Kormanyos2015,Echeverry2016,Slobodeniuk2016,Koperski2017,Molas_2017_DX_BDX,Zhang2017,Molas2019,Robert2020}.
Accordingly, S-TMD MLs are commonly classified as \textit{darkish} or \textit{bright}: in MoS$_2$, WS$_2$, and WSe$_2$, the lowest spin-forbidden excitons lie below their bright counterparts, whereas in MoSe$_2$ and MoTe$_2$ the lowest excitonic state is optically allowed~\cite{Molas_2017_DX_BDX,Zhang2017,Koperski2017,Lu_2020,Robert2020}.
This energetic ordering strongly affects the optical response, as relaxation into low-lying dark states suppresses photoluminescence (PL) and modifies exciton and trion recombination dynamics, whereas the reduced radiative decay of dark states enables the formation of long-lived excitonic and valley reservoirs
~\cite{Courtade2017,Robert2017, Li2018, Chen2018, Barbone2018, Paur2019, Li2019, Lireplica2019,LiuGate2019, Li2019momentum, Molas2019, LiuValley, Liu2020, Robert2020, He2020, Arora2020, Robert2021, Zinkiewicz2021,Zinkiewicz2022}.
In S-TMD MLs, an in-plane magnetic field mixes opposite-spin conduction-band states and redistributes oscillator strength between bright and spin-forbidden excitons, thereby enabling optical access to dark excitonic states and providing insight into the dark–bright splitting and conduction-band spin structure~\cite{Molas_2017_DX_BDX,Zhang2017,Lu_2020,Robert2020,Kipczak2024, krasucki2026dopinginducedbrighteningdarkexcitons}.
Related brightening mechanisms based on symmetry breaking, strain, proximity effects, and nanophotonic coupling have also been explored~\cite{Huo2017, Wiesner2022,Kipczak2025, FerroelectricBrightening2026,PlasmonBrightening2025}.

By contrast, the inverse process, the loss of optical activity of an initially bright excitonic state, here referred to as \textit{darkening}, has remained essentially unexplored in S-TMD MLs.
Although strain- and composition-induced bright-to-dark crossovers have been reported in selected systems~\cite{Wiesner2022,Janus2025}, and enhanced in-plane magnetic responses have recently been observed in MoWSe$_2$ alloys~\cite{Schindler2026}, these studies did not demonstrate a controlled magnetic-field-induced darkening of a bright exciton accompanied by activation of its dark counterpart.
Here we establish this missing complementary regime: magnetic brightening of the dark exciton occurs together with pronounced magnetic darkening of the initially bright state.
Alloying provides the key control parameter because, in Mo$_x$W$_{1-x}$Se$_2$, composition strongly modifies the band-edge spin structure~\cite{OlkowskaPucko2026,Schindler2026} and, through the resulting changes in the low-energy electronic states, can also enhance the effective magnetic coupling between bright and dark excitons.

In this Letter, we investigate the interplay between bright and dark excitons in an hBN-encapsulated Mo$_{0.68}$W$_{0.32}$Se$_2$ ML alloy using low-temperature PL spectroscopy in in-plane magnetic fields $B_\parallel$ up to $25~\mathrm{T}$, complemented by density-functional-theory calculations including $GW$ corrections and a theoretical $\mathbf{k}\cdot\mathbf{p}$ analysis. 
The $\mathbf{k}\cdot\mathbf{p}$ model describes the coupled evolution of the bright and dark excitons under an in-plane magnetic field, while complementary DFT-$GW$ calculations support a spin-forbidden lowest-energy transition for an alloy composition close to that investigated experimentally.
The alloyed ML exhibits a bright, MoSe$_2$-like emission at zero field. 
Surprisingly, an additional emission line emerges below the bright exciton for $B_\parallel\gtrsim15~\mathrm{T}$, with an intensity that increases quadratically with $B_\parallel$, as expected for the magnetic brightening of a spin-forbidden dark exciton. 
Remarkably, the emergence of the dark-exciton emission is accompanied by a pronounced suppression of the initially bright exciton, revealing the simultaneous magnetic brightening of the dark state and darkening of its bright counterpart. 
The magnetic-field evolution of the exciton energies and intensities is consistently captured by the bright-dark exciton model, yielding a zero-field bright-dark splitting of $\Delta=14.2\pm 1.3~\mathrm{meV}$ and an effective in-plane $g$ factor of $|g_\parallel|=5.2\pm 0.2$.

\textit{Theoretical background--} To describe the magnetic-field-induced mixing between bright and dark excitons in S-TMD MLs, we employ a $\mathbf{k}\cdot\mathbf{p}$ approach formulated in the basis of intravalley A-exciton states~\cite{Slobodeniuk2016,Koperski2017,Wang2018}. 
At zero magnetic field, each excitonic state is specified by (i) the valley $\mathbf{K}^{\pm}$ occupied by the electron and hole and (ii) the spin indices $s_v,s_c=\uparrow,\downarrow$ of the valence- and conduction-band states forming the exciton.

The configurations $s_v=s_c=\uparrow$ in the $\mathbf{K}^+$ valley and $s_v=s_c=\downarrow$ in the $\mathbf{K}^-$ valley correspond to the bright A excitons, denoted by $|\mathbf{K}^{\pm},\mathrm{b}\rangle$, respectively. 
The opposite-spin configurations, $s_v=\uparrow$, $s_c=\downarrow$ in $\mathbf{K}^+$ and $s_v=\downarrow$, $s_c=\uparrow$ in $\mathbf{K}^-$, form the dark A-exciton states $|\mathbf{K}^{\pm},\mathrm{d}\rangle$. 
The bright states possess in-plane transition dipole moments~\cite{Slobodeniuk2016,Slobodeniuk2019_BDXT},
$\langle0|\widehat{\mathbf d}|\mathbf K^\pm,\mathrm b\rangle =-id_\parallel(\pm\mathbf e_x+i\mathbf e_y),$
and couple to $\sigma^\pm$ circularly polarized light at normal incidence. 
In contrast, the dark states possess out-of-plane transition dipole moments~\cite{Slobodeniuk2016,Robert2017_DX,Wang2017,Slobodeniuk2019_BDXT},
$\langle0|\widehat{\mathbf d}|\mathbf K^\pm,\mathrm d\rangle =id_\perp\mathbf e_z,$ 
and couple to $z$-polarized light.

An in-plane magnetic field $\mathbf{B}_\parallel=(B_x,B_y)$ applied parallel to the ML's plane mixes opposite-spin CB states within the same valley through the Zeeman interaction $g_\parallel\mu_BB_\pm/2$, where $B_\pm=B_x\pm iB_y$.
In the excitonic basis
$ |\mathbf K^+,\mathrm b\rangle, |\mathbf K^-,\mathrm b\rangle, |\mathbf K^+,\mathrm d\rangle$ and $|\mathbf K^-,\mathrm d\rangle,$ 
the corresponding Hamiltonian reads~\cite{Molas_2017_DX_BDX,Molas2019}
\begin{align}
H=\frac12\left[
          \begin{array}{cccc}
             \Delta & 0 & g_\parallel\mu_BB_- & 0  \\
             0 & \Delta & 0 & g_\parallel\mu_BB_+   \\
						 g_\parallel\mu_B B_+ &  0  & -\Delta + \delta & \delta \\
						 0 &  g_\parallel\mu_B B_-  & \delta & -\Delta + \delta \\
          \end{array}
        \right].
\end{align}
where $\Delta$ denotes the zero-field bright-dark exciton splitting. 
The short-range electron-hole exchange interaction, described by $\delta$, mixes the dark excitons from opposite valleys and splits them into gray and strictly dark states~\cite{Robert2017,Molas2019}. 
Since this splitting is typically much smaller than the bright-dark separation ($\delta\ll\Delta$)~\cite{Robert2017, Molas2019, Robert2020, Zinkiewicz2020, Kapuscinski2021}, we set $\delta=0$ in the following.

The Hamiltonian then separates into two $2\times2$ blocks describing bright-dark mixing independently within each valley:
\begin{equation}
H_\pm=\frac12\left[
          \begin{array}{cc}
            \Delta  & g_\parallel\mu_BB_\mp \\
            g_\parallel\mu_B B_\pm &  -\Delta  \\
          \end{array}
        \right].
\end{equation}


Writing $B_x=B_\parallel\cos\phi$ and $B_y=B_\parallel\sin\phi$, such that $B_\pm=B_\parallel e^{\pm i\phi}$, and introducing the mixing angle
$\tan\vartheta=\frac{g_\parallel\mu_BB_\parallel}{\Delta}$, we obtain
\begin{equation}
H_\pm=\frac12 \sqrt{\Delta^2+(g_\parallel\mu_BB_\parallel)^2}\left[
          \begin{array}{cc}
            \cos\vartheta  & e^{\mp i\phi}\sin\vartheta \\
             e^{\pm i\phi}\sin\vartheta &  -\cos\vartheta \\
          \end{array}
        \right].
\end{equation}

The resulting bright-like and dark-like branches remain valley degenerate and have energies
\begin{align}
\label{r:energy}
E_{\mathrm{B,D}}=\pm \frac{1}{2}\sqrt{\Delta^2+(g_\parallel\mu_BB_\parallel)^2}.
\end{align}
The corresponding eigenstates are
\begin{align}
|\text{B}_\pm\rangle=&\cos\Big(\frac{\vartheta}{2}\Big)|\mathbf{K}^\pm,\mathrm{b}\rangle + 
e^{\pm i\phi}\sin\Big(\frac{\vartheta}{2}\Big) |\mathbf{K}^\pm,\mathrm{d}\rangle, \\
|\text{D}_\pm\rangle=& -e^{\mp i\phi}\sin\Big(\frac{\vartheta}{2}\Big) |\mathbf{K}^\pm,\mathrm{b}\rangle +
\cos\Big(\frac{\vartheta}{2}\Big)|\mathbf{K}^\pm,\mathrm{d}\rangle.
\end{align} 

The bright-state admixture in the dark-like exciton is therefore~\cite{Molas_2017_DX_BDX}
\begin{equation}
w(B_\parallel)=
\sin^2\left(\frac{\vartheta}{2}\right).
\label{eq:bright_admixture}
\end{equation}
Since $\tan\vartheta=g_\parallel\mu_\mathrm{B}B_\parallel/\Delta$, the leading contribution to $w$ scales as $B_\parallel^2$. 
Within the standard description of magnetic brightening of dark excitons~\cite{Molas_2017_DX_BDX}, the intensity of the dark-exciton emission is expected to be proportional to the bright-exciton admixture $w$, the dark-exciton population $n_\mathrm{D}$, and the zero-field bright-exciton intensity $I_\mathrm{B}$:
\begin{equation}
I_\mathrm{D}\propto n_\mathrm{D} I_\mathrm{B} w
\propto n_\mathrm{D} I_\mathrm{B} B_\parallel^2 .
\label{eq:ID_quadratic}
\end{equation}

The same mixing also modifies the bright-like exciton. 
As discussed in Sec.~S1 of the Supplemental Material (SM), the change in the collected emission arising solely from the field-induced modification of the bright-exciton transition dipole is small under our experimental conditions and cannot account for the pronounced quenching of the A-exciton emission discussed below.
We therefore consider the magnetic-field dependence of the bright-exciton population.

In a minimal model, we assume thermal equilibrium at temperature $T$ and take the population of the bright-like branch to follow the Boltzmann distribution,
$n_\mathrm{B}(B_\parallel)\propto 
\exp\left[-\frac{E_\mathrm{B}(B_\parallel)}{k_\mathrm{B}T}\right].
$
Using the field dependence of the bright-like exciton energy in Eq.~(\ref{r:energy}), the population normalized to its zero-field value becomes
\begin{align}
\label{eq:population}
\frac{n_\text{B}(B_\parallel)}{n_\text{B}(0)}=\exp\Big(-\frac{\Delta}{2k_\text{B}T}\Big[\sqrt{1+\left(\frac{g_\parallel\mu_BB_\parallel}{\Delta}\right)^2}-1\Big]\Big).
\end{align}

\begin{figure}[t]
	\centering
	\includegraphics[width=1\linewidth]{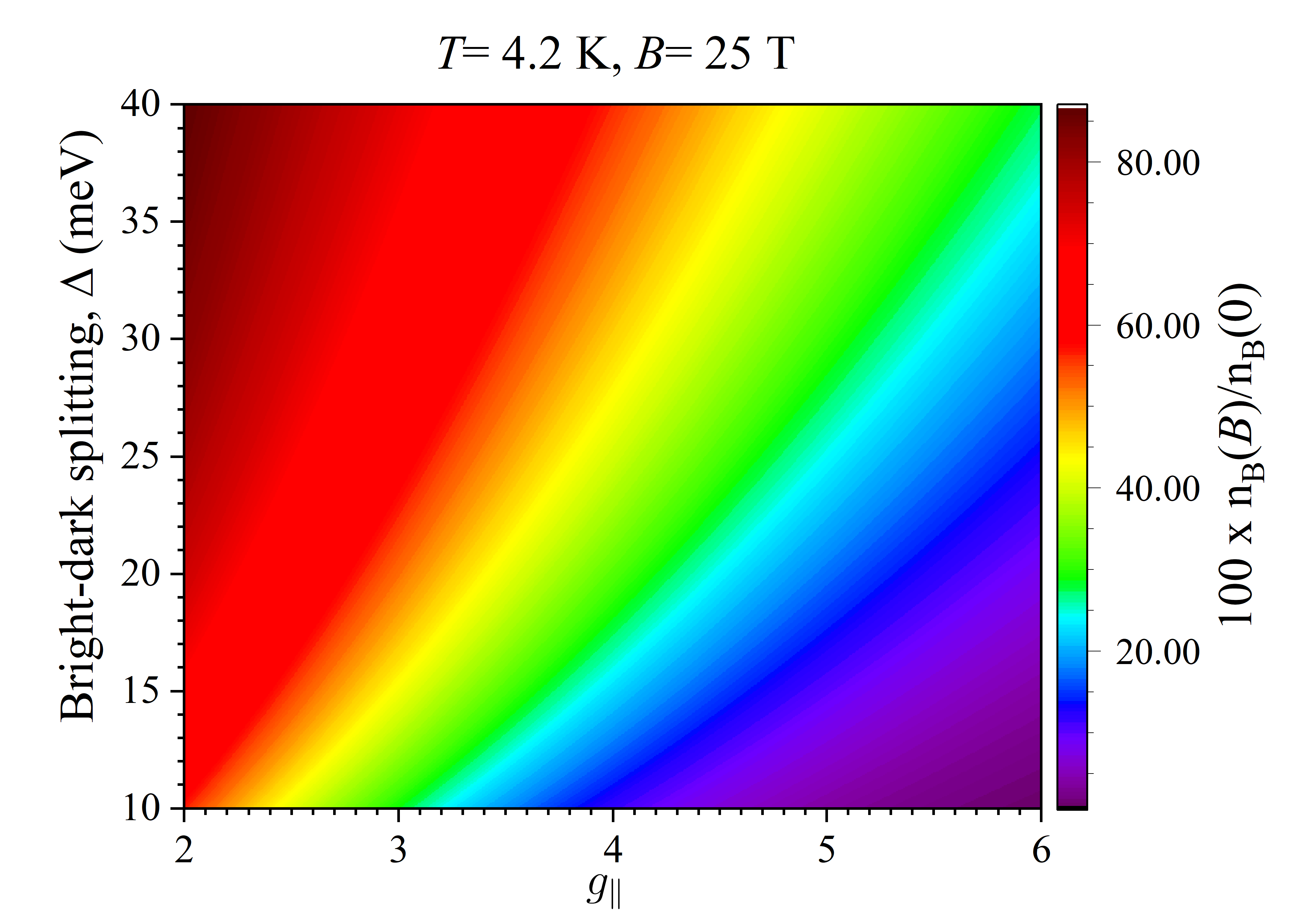}
	\caption{False-color map of the calculated normalized bright-exciton population $n_\text{B}(B_\parallel)/n_\text{B}(0)$, expressed as a percentage at $T=4.2$~K and $B_\parallel=25$~T as a function of the bright-dark exciton splitting $\Delta$ and the in-plane coupling parameter $g_\parallel$ using Eq.~(\ref{eq:population}).}
	\label{fig:population}
\end{figure}

This expression qualitatively describes the magnetic-field-induced evolution of the bright-exciton population $n_\mathrm{B}$. 
To illustrate the roles of the bright-dark exciton splitting $\Delta$ and the in-plane coupling parameter $g_\parallel$, Fig.~\ref{fig:population} shows the calculated $n_\text{B}(B_\parallel)/n_\text{B}(0)$ at $T=4.2$~K and $B_\parallel=25$~T using Eq.~(\ref{eq:population}).
The strongest field-induced depletion of the bright-exciton population is expected for small $\Delta$ and large $g_\parallel$.

For WS$_2$ and WSe$_2$ MLs, where the bright-dark exciton splitting is typically $\Delta\approx40$~meV~\cite{Wang2017,Zhou2017,Molas2019,Robert2020,Zinkiewicz2020} and $g_\parallel\approx2$~\cite{Molas2019,Robert2020}, the predicted reduction of the bright-exciton population at $25$~T is only about 10\%. 
Such a weak effect may explain why a pronounced magnetic-field-induced depletion of the bright-exciton emission has not been reported in previous studies of dark-exciton brightening~\cite{Molas_2017_DX_BDX,Zhang2017,Zinkiewicz2020,Robert2020}.
This establishes a simple criterion for strong magnetic darkening: a small bright-dark splitting combined with strong in-plane coupling.

\begin{figure}[t]
	\centering
	\includegraphics[width=1\linewidth]{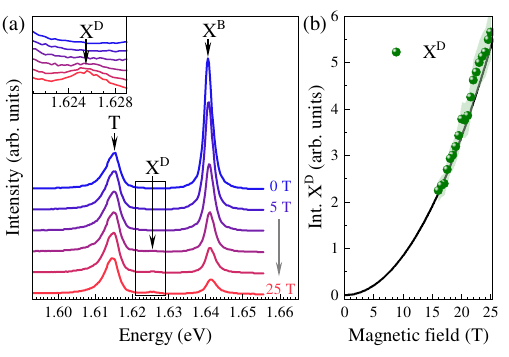}
	\caption{(a) Evolution of the low-temperature photoluminescence spectra of the Mo$_{0.68}$W$_{0.32}$Se$_2$ ML measured under an in-plane magnetic field up to 25~T.
    The excitation energy and laser power were 2.41~eV and 10~$\mu$W, respectively.
    (b) Integrated intensity of the X$^\mathrm{D}$ line as a function of $B_\parallel$, together with the quadratic fit described in the text.
    The shaded area represents the uncertainty of the extracted integrated intensity.
    }
	\label{fig:pl}
\end{figure}

\textit{Magnetic brightening of the dark exciton--} To test this prediction, we performed low-temperature PL measurements on an hBN-encapsulated Mo$_{0.68}$W$_{0.32}$Se$_2$ ML in the Voigt geometry, with an in-plane magnetic field $B_\parallel$ up to $25~\mathrm{T}$. 
Details of the sample fabrication and magneto-optical measurements are provided in the Methods section of the SM~\cite{SM}.
The resulting magnetic-field evolution of the PL spectra is shown in Fig.~\ref{fig:pl}(a).
The same data plotted on a logarithmic intensity scale are presented in Sec.~S2 of the SM~\cite{SM}, where the weak X$^\mathrm{D}$ emission is more clearly visible.

At zero magnetic field, the spectrum closely resembles that of bright MoSe$_2$~\cite{Robert2017,Molas2019Energy,Koperski2019,Lu_2020,Robert2020,Oreszczuk2023}. 
Following the assignment established for MoWSe$_2$ alloy MLs, and in particular for Mo$_{0.68}$W$_{0.32}$Se$_2$~\cite{OlkowskaPucko2026}, the X$^\mathrm{B}$ and T lines are attributed to the bright neutral and charged excitons, respectively.

For $B_\parallel>15~\mathrm{T}$, an additional emission line, labeled X$^\mathrm{D}$, emerges on the low-energy side of X$^\mathrm{B}$, separated from it by approximately $15~\mathrm{meV}$ [inset of Fig.~\ref{fig:pl}(a)]. 
Its integrated intensity, extracted from Lorentzian fits, increases with $B_\parallel$ [Fig.~\ref{fig:pl}(b)]. 
The field dependence of the X$^\mathrm{D}$ intensity is well described by the quadratic relation $I_{\mathrm{X}^{\mathrm{D}}}(B_\parallel)=\alpha B_\parallel^2$,
yielding $\alpha=(7.51\pm0.15)\times10^{-3}~\mathrm{T}^{-2}$. 
This quadratic dependence is consistent with Eq.~(\ref{eq:ID_quadratic}), according to which the dark-exciton emission intensity is governed by the bright-state admixture induced by the in-plane magnetic field. 
Such a $B_\parallel^2$ scaling is a characteristic signature of the magnetic brightening of spin-forbidden dark excitons and has been previously observed in S-TMD MLs~\cite{Molas_2017_DX_BDX,Zhang2017,Molas2019,Robert2020,Zinkiewicz2020}.

The observed X$^\mathrm{B}$-X$^\mathrm{D}$ separation of approximately 15~meV is close to the bright-dark splitting reported for MoS$_2$ MLs (around 15~meV)~\cite{Robert2020}, and substantially smaller than the values of about 40~meV typical of WS$_2$ and WSe$_2$~\cite{Robert2017_DX, Molas2019, Zinkiewicz2020, Liu2020,He2020,Zinkiewicz2021, Kipczak2024}.
The assignment of X$^\mathrm{D}$ to the neutral dark exciton is further supported by spin-resolved band-structure calculations, which are consistent with a spin-forbidden lowest-energy transition at the $\mathbf{K}$ valleys; see Sec.~S3 of the SM~\cite{SM}.

\begin{figure}[t]
	\centering
	\includegraphics[width=1\linewidth]{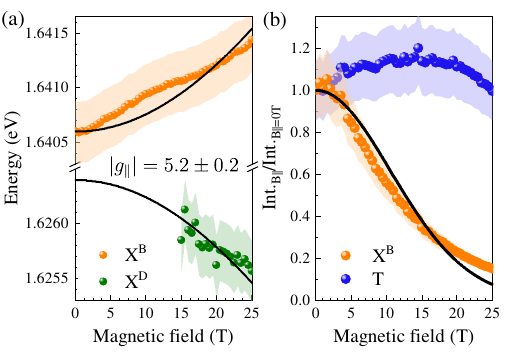}
    \caption{(a) Energies of the bright X$^\mathrm{B}$ and dark X$^\mathrm{D}$ excitons as a function of the in-plane magnetic field $B_\parallel$.
    The solid lines are fits based on the bright-dark exciton mixing model described by Eq.~(\ref{r:g_inplane}).
    The fit was performed over the full available field range for X$^\mathrm{B}$, 0-25~T, and over the range where X$^\mathrm{D}$ is resolved, 15-25~T.
    The shaded areas indicate the uncertainties of the extracted exciton energies.
    (b) Magnetic-field dependence of the integrated intensities of the X$^\mathrm{B}$ and T emission lines, normalized to their zero-field values.
    The shaded areas indicate the uncertainties of the extracted integrated intensities.
    The solid line shows the population model described by Eq.~(\ref{eq:population}). 
    }
	\label{fig:g_factor}
\end{figure}

\textit{Magnetic-field evolution of bright and dark excitons--} Additionally, a striking feature of the spectra is the simultaneous suppression of the X$^\mathrm{B}$ emission with increasing $B_\parallel$, while the T intensity remains nearly unchanged over the entire magnetic-field range [see Fig.~\ref{fig:g_factor}(b)]. 
This behavior goes beyond the conventional magnetic brightening of dark excitons observed in binary S-TMD MLs, where the X$^\mathrm{B}$ emission remains essentially unaffected by the applied in-plane magnetic field~\cite{Molas_2017_DX_BDX,Zhang2017,Molas2019,Robert2020,Zinkiewicz2020}. 
Thus, in the studied Mo$_{0.68}$W$_{0.32}$Se$_2$ alloy ML, the magnetic brightening of X$^\mathrm{D}$ is accompanied by a pronounced magnetic-field-induced darkening of X$^\mathrm{B}$.

To analyze this effect quantitatively, we first extract the magnetic-field evolution of the X$^\mathrm{B}$ and X$^\mathrm{D}$ energies from the measured PL spectra, Fig.~\ref{fig:g_factor}(a). 
With increasing $B_\parallel$, the X$^\mathrm{B}$ and X$^\mathrm{D}$ lines exhibit characteristic blue- and redshifts, respectively, resulting from the bright-dark exciton mixing.

Following the two-level bright-dark mixing model introduced above, the field evolution of the bright-like and dark-like exciton branches is described by
\begin{align}
E_{\mathrm{B,D}}(B_\parallel)=E_0\pm\frac{1}{2}\sqrt{\Delta^2+\left(g_\parallel\mu_\mathrm{B}B_\parallel\right)^2
},
\label{r:g_inplane}
\end{align}
where the $+$ and $-$ signs correspond to the X$^\mathrm{B}$ and X$^\mathrm{D}$ branches, respectively. 
Here, $E_0$ denotes the mean energy of the bright and dark exciton states, whereas $\Delta$ is their zero-field energy splitting. 
These quantities are defined as
\begin{align}
E_0 &= \frac{E_\mathrm{B}(B_\parallel)+E_\mathrm{D}(B_\parallel)}{2},& \Delta &= E_\mathrm{B}(0)-E_\mathrm{D}(0),
\label{eq:E0_Delta_def}
\end{align}
where $E_\mathrm{B}(B_\parallel)$ and $E_\mathrm{D}(B_\parallel)$ denote the energies of the bright-like and dark-like exciton branches at a given magnetic field, respectively.

Since the dark exciton is not optically resolved at zero magnetic field, $E_\mathrm{D}(0)$ cannot be determined directly from the PL spectrum. 
We therefore adopt the following procedure. 
First, $E_0$ is determined by averaging the quantity $[E_\mathrm{B}(B_\parallel)+E_\mathrm{D}(B_\parallel)]/2$ over the experimental data in the field range $15$-$25$~T, where both excitonic transitions are resolved. 
The zero-field bright-dark splitting is then obtained from $\Delta=2[E_\mathrm{B}(0)-E_0]$.
The field dependence of the X$^\mathrm B$ and X$^\mathrm D$ energies [Fig.~\ref{fig:g_factor}(a)] is then fitted using Eq.~(\ref{r:g_inplane}), with $E_0=1.63350\pm 0.00040~\mathrm{eV}$ and $\Delta=14.2\pm 1.3~\mathrm{meV}$ fixed to the values determined above.

Thus, $|g_\parallel|$ is the only free fitting parameter.
The resulting curves, shown as solid lines in Fig.~\ref{fig:g_factor}(a), reproduce the experimental data very well.
The extracted value, $|g_{\parallel}|=5.2\pm0.2$, is substantially larger than the value of approximately 2 reported for binary S-TMD MLs~\cite{Molas2019,Robert2020}.
This enhanced magnetic response can be qualitatively understood as a consequence of alloy-induced modifications of the band-edge electronic structure, consistent with the strong composition dependence of the out-of-plane excitonic $g_\perp$ factor recently observed in Mo$_x$W$_{1-x}$Se$_2$ alloys~\cite{OlkowskaPucko2026}.
Changes in the relative positions of the bands modify the contributions of remote bands to the low-energy states and can therefore substantially renormalize the effective in-plane coupling $g_{\parallel}$ governing the bright-dark mixing.
A larger $|g_{\parallel}|$, together with the reduced bright-dark splitting near the conduction-band spin-orbit crossover, favors particularly strong magnetic mixing in the alloy.
Interestingly, an effective in-plane excitonic $g_{\parallel}$ factor of $6.4\pm0.3$ was recently reported for a Mo$_{0.7}$W$_{0.3}$Se$_2$ ML~\cite{Schindler2026}, further highlighting the unusually strong in-plane magnetic response of Mo$_x$W$_{1-x}$Se$_2$ alloys in this composition range.
A more detailed effective-medium description is provided in Sec.~S4 of the SM~\cite{SM}.

We note that the extracted value of $|g_\parallel|$ is sensitive to the uncertainties in $E_0$ and $\Delta$.
In particular, the determination of $|g_\parallel|$ strongly depends on the assumed value of $\Delta$.
A detailed analysis of this sensitivity is provided in Sec.~S5 of the SM~\cite{SM}.

We now turn to the magnetic-field dependence of the X$^\mathrm{B}$ emission intensity. 
A pronounced suppression of X$^\mathrm{B}$ is already evident in the PL spectra shown in Fig.~\ref{fig:pl}(a). 
To quantify this effect, we extract the integrated intensities of the X$^\mathrm{B}$ and T emission lines from Lorentzian fits and normalize them to their zero-field values [Fig.~\ref{fig:g_factor}(b)].

While the T intensity remains nearly constant over the entire magnetic-field range, the X$^\mathrm{B}$ emission is strongly suppressed, decreasing by approximately $85\%$ at $B_\parallel=25~\mathrm{T}$. 
The distinct field evolution of the two transitions rules out a trivial loss of the PL signal, for example, due to a drift of the excitation or collection spot across an inhomogeneous sample. 
Instead, it indicates that the in-plane magnetic field selectively suppresses the neutral bright-exciton emission.

This magnetic-field-induced darkening of X$^\mathrm{B}$ can be understood within the population model introduced above.
Assuming thermal equilibrium at $T=4.2~\mathrm{K}$, the relative population of the bright-like exciton branch is given by Eq.~(\ref{eq:population}). 
Using the experimentally determined parameters $\Delta=14.2\pm 1.3~\mathrm{meV}$ and $|g_\parallel|=5.2\pm 0.2$, we obtain $n_\mathrm{B}(25~\mathrm{T})/n_\mathrm{B}(0)\simeq0.08\pm 0.02$, corresponding to a $92\%$ reduction of the bright-exciton population. 
This value is close to the experimentally observed suppression of the X$^\mathrm{B}$ emission by approximately $85\%$.

Using the same values of $\Delta$ and $|g_\parallel|$, we calculate the full magnetic-field dependence of $n_\mathrm{B}(B_\parallel)/n_\mathrm{B}(0)$ from Eq.~(\ref{eq:population}). 
The resulting curve, shown as the solid line in Fig.~\ref{fig:g_factor}(b), closely reproduces the measured evolution of the normalized X$^\mathrm{B}$ intensity. This agreement supports the interpretation that the pronounced magnetic darkening of the bright exciton is dominated by the field-induced increase in the energy of the bright-like branch and the consequent depletion of its thermal population.

\textit{Summary--} We have demonstrated that an in-plane magnetic field can simultaneously brighten a spin-forbidden dark exciton and strongly suppress its bright counterpart in a Mo$_{0.68}$W$_{0.32}$Se$_2$ ML alloy. 
Although the zero-field PL resembles that of a bright MoSe$_2$-like ML, the high-field response reveals a dark neutral exciton below the bright state. 
The coupled evolution of the two exciton branches is captured by a bright-dark mixing model, while the pronounced quenching of the bright-exciton emission is accounted for by the field-induced increase in the energy of the bright-like branch and the consequent depletion of its thermal population. 
The relatively small bright-dark splitting and enhanced in-plane coupling place the alloy in a regime where magnetic darkening becomes directly observable.
More generally, because the effect is controlled by the competition between $\Delta$ and $g_\parallel\mu_\mathrm{B}B_\parallel$, it should be enhanced in excitonic systems with near-degenerate bright and dark states.
These findings establish composition as a route to engineering the bright-dark exciton hierarchy and to magnetic control of exciton populations and light emission in two-dimensional semiconductors.

\section{Acknowledgements}
The work was supported by the National Science Centre, Poland (grants no. 2022/45/N/ST3/01887 and 2025/57/B/ST5/03288). 
The publication was created as part of a project co-financed by the Polish Ministry of Science and Higher Education under contract no. 2025/WK/01. 
We acknowledge the support of the LNCMI-CNRS, a member of the European Magnetic Field Laboratory (EMFL). 
A.O.S. acknowledges the support of the Czech Science Foundation (project GA\v{C}R 26-21965S).
A.O.S. and M.P. acknowledge support from the CENTERA2, FENG.02.01-IP.05-5 T004/23 project funded within the IRA program of the FNP Poland, co-financed by the EU FENG Programme.
A.P. and E.B. acknowledge funding by the European Union's EIC Pathfinder Open 2024 under grant agreement 101186642 (VALERIA).
E.B. acknowledges support from the Italian Science Fund (Fondo Italiano per la Scienza, FIS), FIS 3 Call (Project INCOGNITO, ID: FIS-2024-07158).
K.W. and T.T. acknowledge support from the CREST (JPMJCR24A5), JST and World Premier International Research Center Initiative (WPI), MEXT, Japan.

\bibliographystyle{apsrev4-2}
\bibliography{biblio}

@article{Oreszczuk2023,
doi = {10.1088/2053-1583/acefe3},
year = {2023},
month = {aug},
publisher = {IOP Publishing},
volume = {10},
number = {4},
pages = {045019},
author = {K Oreszczuk and A Rodek and M Goryca and T Kazimierczuk and M Raczyński and J Howarth and T Taniguchi and K Watanabe and M Potemski and P Kossacki},
title = {Enhancement of electron magnetic susceptibility due to many-body interactions in monolayer ${\textit{\mathrm{MoSe}}_{2}$}},
journal = {2D Materials}
}

@article{Zhang2017,
author={Zhang, Xiao-Xiao and Cao, Ting and Lu, Zhengguang and Lin, Yu-Chuan and Zhang, Fan and Wang, Ying and Li, Zhiqiang and Hone, James C. and Robinson, Joshua A. and Smirnov, Dmitry and Louie, Steven G. and Heinz, Tony F.},
title={Magnetic brightening and control of dark excitons in monolayer WSe$_2$},
journal={Nature Nanotechnology},
year={2017},
month={Sep},
day={01},
volume={12},
number={9},
pages={883-888},
doi={10.1038/nnano.2017.105}
}

@article{Robert2017,
  title = {Fine structure and lifetime of dark excitons in transition metal dichalcogenide monolayers},
  author = {Robert, C. and Amand, T. and Cadiz, F. and Lagarde, D. and Courtade, E. and Manca, M. and Taniguchi, T. and Watanabe, K. and Urbaszek, B. and Marie, X.},
  journal = {Phys. Rev. B},
  volume = {96},
  issue = {15},
  pages = {155423},
  numpages = {8},
  year = {2017},
  month = {Oct},
  publisher = {American Physical Society},
  doi = {10.1103/PhysRevB.96.155423},
}

@article{Wang2017,
  title = {In-Plane Propagation of Light in Transition Metal Dichalcogenide Monolayers: Optical Selection Rules},
  author = {Wang, G. and Robert, C. and Glazov, M. M. and Cadiz, F. and Courtade, E. and Amand, T. and Lagarde, D. and Taniguchi, T. and Watanabe, K. and Urbaszek, B. and Marie, X.},
  journal = {Phys. Rev. Lett.},
  volume = {119},
  issue = {4},
  pages = {047401},
  numpages = {7},
  year = {2017},
  month = {Jul},
  publisher = {American Physical Society},
  doi = {10.1103/PhysRevLett.119.047401}
}

@article{Paur2019,
author={Paur, Matthias and Molina-Mendoza, Aday J. and Bratschitsch, Rudolf and Watanabe, Kenji and Taniguchi, Takashi and Mueller, Thomas},
title={Electroluminescence from multi-particle exciton complexes in transition metal dichalcogenide semiconductors},
journal={Nature Communications},
year={2019},
month={Apr},
day={12},
volume={10},
number={1},
pages={1709},
doi={10.1038/s41467-019-09781-y}
}

@Article{Barbone2018,
author={Barbone, Matteo and Montblanch, Alejandro R.-P. and Kara, Dhiren M. and Palacios-Berraquero, Carmen and Cadore, Alisson R.
and De Fazio, Domenico and Pingault, Benjamin and Mostaani, Elaheh
and Li, Han and Chen, Bin and Watanabe, Kenji and Taniguchi, Takashi
and Tongay, Sefaattin and Wang, Gang and Ferrari, Andrea C.
and Atat{\"u}re, Mete},
title={Charge-tuneable biexciton complexes in monolayer WSe$_2$},
journal={Nature Communications},
year={2018},
month={Sep},
day={13},
volume={9},
number={1},
pages={3721},
doi={10.1038/s41467-018-05632-4}
}

@Article{Chen2018,
author={Chen, Shao-Yu and Goldstein, Thomas and Taniguchi, Takashi
and Watanabe, Kenji and Yan, Jun},
title={Coulomb-bound four- and five-particle intervalley states in an atomically-thin semiconductor},
journal={Nature Communications},
year={2018},
month={Sep},
day={13},
volume={9},
number={1},
pages={3717},
doi={10.1038/s41467-018-05558-x}
}

@article{Liu2020,
  title = {Multipath Optical Recombination of Intervalley Dark Excitons and Trions in Monolayer WSe$_{2}$},
  author = {Liu, Erfu and van Baren, Jeremiah and Liang, Ching-Tarng and Taniguchi, Takashi and Watanabe, Kenji and Gabor, Nathaniel M. and Chang, Yia-Chung and Lui, Chun Hung},
  journal = {Phys. Rev. Lett.},
  volume = {124},
  issue = {19},
  pages = {196802},
  numpages = {7},
  year = {2020},
  month = {May},
  publisher = {American Physical Society},
  doi = {10.1103/PhysRevLett.124.196802}
}

@article{Molas2019,
  title = {Probing and Manipulating Valley Coherence of Dark Excitons in Monolayer ${\mathrm{WSe}}_{2}$},
  author = {Molas, M. R. and Slobodeniuk, A. O. and Kazimierczuk, T. and Nogajewski, K. and Bartos, M. and Kapu\ifmmode \acute{s}\else \'{s}\fi{}ci\ifmmode \acute{n}\else \'{n}\fi{}ski, P. and Oreszczuk, K. and Watanabe, K. and Taniguchi, T. and Faugeras, C. and Kossacki, P. and Basko, D. M. and Potemski, M.},
  journal = {Phys. Rev. Lett.},
  volume = {123},
  issue = {9},
  pages = {096803},
  numpages = {6},
  year = {2019},
  month = {Aug},
  publisher = {American Physical Society},
  doi = {10.1103/PhysRevLett.123.096803}
}

@article{He2020,
    author={He, Minhao and Rivera, Pasqual and Van Tuan, Dinh and Wilson, Nathan P. and Yang, Min and Taniguchi, Takashi and Watanabe, Kenji and Yan, Jiaqiang and Mandrus, David G. and Yu, Hongyi and Dery, Hanan and Yao, Wang and Xu, Xiaodong},
    title={Valley phonons and exciton complexes in a monolayer semiconductor},
    journal={Nature Communications},
    year={2020},
    month={Jan},
    day={30},
    volume={11},
    number={1},
    pages={618},
    issn={2041-1723},
    doi={10.1038/s41467-020-14472-0}
}

@article{LiuValley,
  title = {Valley-selective chiral phonon replicas of dark excitons and trions in monolayer $\mathrm{WS}{\mathrm{e}}_{2}$},
  author = {Liu, Erfu and van Baren, Jeremiah and Taniguchi, Takashi and Watanabe, Kenji and Chang, Yia-Chung and Lui, Chun Hung},
  journal = {Phys. Rev. Research},
  volume = {1},
  issue = {3},
  pages = {032007},
  numpages = {7},
  year = {2019},
  month = {Oct},
  publisher = {American Physical Society},
  doi = {10.1103/PhysRevResearch.1.032007}
}

@article{LiuGate2019,
  title = {Gate Tunable Dark Trions in Monolayer WSe$_{2}$},
  author = {Liu, Erfu and van Baren, Jeremiah and Lu, Zhengguang and Altaiary, Mashael M. and Taniguchi, Takashi and Watanabe, Kenji and Smirnov, Dmitry and Lui, Chun Hung},
  journal = {Phys. Rev. Lett.},
  volume = {123},
  issue = {2},
  pages = {027401},
  numpages = {6},
  year = {2019},
  month = {Jul},
  publisher = {American Physical Society},
  doi = {10.1103/PhysRevLett.123.027401}
}

@Article{Li2019,
author = {Li, Zhipeng and Wang, Tianmeng and Lu, Zhengguang and Khatoniar, Mandeep and Lian, Zhen and Meng, Yuze and Blei, Mark and Taniguchi, Takashi and Watanabe, Kenji and McGill, Stephen A. and Tongay, Sefaattin and Menon, Vinod M. and Smirnov, Dmitry and Shi, Su-Fei},
title = {Direct Observation of Gate-Tunable Dark Trions in Monolayer WSe$_2$},
journal = {Nano Letters},
volume = {19},
number = {10},
pages = {6886-6893},
year = {2019},
doi = {10.1021/acs.nanolett.9b02132}
}

@Article{Lireplica2019,
author={Li, Zhipeng and Wang, Tianmeng and Jin, Chenhao and Lu, Zhengguang and Lian, Zhen and Meng, Yuze and Blei, Mark and Gao, Shiyuan and Taniguchi, Takashi and Watanabe, Kenji and Ren, Tianhui and Tongay, Sefaattin and Yang, Li and Smirnov, Dmitry and Cao, Ting
and Shi, Su-Fei},
title={Emerging photoluminescence from the dark-exciton phonon replica in monolayer WSe$_2$},
journal={Nature Communications},
year={2019},
month={Jun},
day={06},
volume={10},
number={1},
pages={2469},
issn={2041-1723},
doi={10.1038/s41467-019-10477-6}
}

@Article{Li2019momentum,
author = {Li, Zhipeng and Wang, Tianmeng and Jin, Chenhao and Lu, Zhengguang and Lian, Zhen and Meng, Yuze and Blei, Mark and Gao, Mengnan and Taniguchi, Takashi and Watanabe, Kenji and Ren, Tianhui and Cao, Ting and Tongay, Sefaattin and Smirnov, Dmitry and Zhang, Lifa and Shi, Su-Fei},
title = {Momentum-Dark Intervalley Exciton in Monolayer Tungsten Diselenide Brightened via Chiral Phonon},
journal = {ACS Nano},
volume = {13},
number = {12},
pages = {14107-14113},
year = {2019},
doi = {10.1021/acsnano.9b06682}
}

@article{Koperski2017,
author = {Koperski, M. and Molas, M. R. and Arora, A. and Nogajewski, K. and Slobodeniuk, A. O. and Faugeras, C. and Potemski, M.},
title = {Optical properties of atomically thin transition metal dichalcogenides: observations and puzzles},
journal = {Nanophotonics},
year = {2017},
volume = {6},
number = {6},
pages = {1289},
url = {https://onlinelibrary.wiley.com/doi/abs/10.1515/nanoph-2016-0165},
doi = {10.1515/nanoph-2016-0165}
}

@article{Koperski2019,
	author={Maciej Koperski and Maciej R Molas and Ashish Arora and Karol Nogajewski and Miroslav Bartos and Jan Wyzula and Diana
	Vaclavkova and Piotr Kossacki and Marek Potemski},
	title={Orbital, spin and valley contributions to Zeeman splitting of excitonic resonances in $\textrm{MoSe}_2$, $\textrm{WSe}_2$ and $\textrm{WS}_2$ Monolayers},
	journal={2D Materials},
	volume={6},
	number={1},
	pages={015001},
	year={2019},
	doi={10.1088/2053-1583/aae14b}
}

@article{Courtade2017,
    title = {{Charged excitons in monolayer ${\mathrm{WSe}}_{2}$: Experiment and theory}},
    author = {Courtade, E. and Semina, M. and Manca, M. and Glazov, M. M. and Robert, C. and Cadiz, F. and Wang, G. and Taniguchi, T. and Watanabe, K. and Pierre, M. and Escoffier, W. and Ivchenko, E. L. and Renucci, P. and Marie, X. and Amand, T. and Urbaszek, B.},
    journal = {Phys. Rev. B},
    volume = {96},
    issue = {8},
    pages = {085302},
    numpages = {12},
    year = {2017},
    month = {Aug},
    publisher = {American Physical Society},
    doi = {10.1103/PhysRevB.96.085302}
}

@Article{Zinkiewicz2020,
author ="Zinkiewicz, M. and Slobodeniuk, A. O. and Kazimierczuk, T. and Kapuściński, P. and Oreszczuk, K. and Grzeszczyk, M. and Bartos, M. and Nogajewski, K. and Watanabe, K. and Taniguchi, T. and Faugeras, C. and Kossacki, P. and Potemski, M. and Babiński, A. and Molas, M. R.",
title  ="Neutral and charged dark excitons in monolayer WS$_2$",
journal  ="Nanoscale",
year  ="2020",
volume  ="12",
issue  ="35",
pages  ="18153-18159",
publisher  ="The Royal Society of Chemistry",
doi  ="10.1039/D0NR04243A"
}

@article{Zinkiewicz2021,
author = {Zinkiewicz, Małgorzata and Woźniak, Tomasz and Kazimierczuk, Tomasz and Kapuscinski, Piotr and Oreszczuk, Kacper and Grzeszczyk, Magdalena and Bartoš, Miroslav and Nogajewski, Karol and Watanabe, Kenji and Taniguchi, Takashi and Faugeras, Clement and Kossacki, Piotr and Potemski, Marek and Babiński, Adam and Molas, Maciej R.},
title = {Excitonic Complexes in n-Doped WS$_2$ Monolayer},
journal = {Nano Letters},
volume = {21},
number = {6},
pages = {2519-2525},
year = {2021},
doi = {10.1021/acs.nanolett.0c05021}
}

@article{Kapuscinski2021,
author={Kapu{\'{s}}ci{\'{n}}ski, Piotr and Delhomme, Alex and Vaclavkova, Diana and Slobodeniuk, Artur O. and Grzeszczyk, Magdalena
and Bartos, Miroslav and Watanabe, Kenji
and Taniguchi, Takashi and Faugeras, Cl{\'e}ment and Potemski, Marek},
title={Rydberg series of dark excitons and the conduction band spin-orbit splitting in monolayer WSe$_2$},
journal={Communications Physics},
year={2021},
month={Aug},
day={19},
volume={4},
number={1},
pages={186},
doi={10.1038/s42005-021-00692-3}
}

@article{Wang2018,
  title = {{Colloquium: Excitons in atomically thin transition metal dichalcogenides}},
  author = {Wang, Gang and Chernikov, Alexey and Glazov, Mikhail M. and Heinz, Tony F. and Marie, Xavier and Amand, Thierry and Urbaszek, Bernhard},
  journal = {Rev. Mod. Phys.},
  volume = {90},
  issue = {2},
  pages = {021001},
  numpages = {25},
  year = {2018},
  month = {Apr},
  publisher = {American Physical Society},
    url = {https://link.aps.org/doi/10.1103/RevModPhys.90.021001},
  doi = {10.1103/RevModPhys.90.021001}
}

@article{Molas2019Energy,
  title = {Energy Spectrum of Two-Dimensional Excitons in a Nonuniform Dielectric Medium},
  author = {Molas, M. R. and Slobodeniuk, A. O. and Nogajewski, K. and Bartos, M. and Bala, \L{}. and Babi\ifmmode \acute{n}\else \'{n}\fi{}ski, A. and Watanabe, K. and Taniguchi, T. and Faugeras, C. and Potemski, M.},
  journal = {Phys. Rev. Lett.},
  volume = {123},
  issue = {13},
  pages = {136801},
  numpages = {6},
  year = {2019},
  month = {Sep},
  publisher = {American Physical Society},
  doi = {10.1103/PhysRevLett.123.136801}
}

@article{Zinkiewicz2022,
    author = {Zinkiewicz, Malgorzata and Grzeszczyk, Magdalena and Kipczak, Lucja and Kazimierczuk, Tomasz and Watanabe, Kenji and Taniguchi, Takashi and Kossacki, Piotr and Babinski, Adam and Molas, Maciej R.},
    title = {The effect of dielectric environment on the brightening of neutral and charged dark excitons in WSe$_2$ monolayer},
    journal = {Applied Physics Letters},
    volume = {120},
    number = {16},
    pages = {163101},
    year = {2022},
    month = {04},
    issn = {0003-6951},
    doi = {10.1063/5.0085950}
  }

@article{Pucko_2023,
doi = {10.1088/2053-1583/aca915},
year = {2022},
month = {dec},
publisher = {IOP Publishing},
volume = {10},
number = {1},
pages = {015018},
author = {Katarzyna Olkowska Pucko and Elena Blundo and Natalia Zawadzka and Salvatore Cianci and Diana Vaclavkova and Piotr Kapuściński and Dipankar Jana and Giorgio Pettinari and Marco Felici and Karol Nogajewski and Miroslav Bartoš and Kenji Watanabe and Takashi Taniguchi and Clement Faugeras and Marek Potemski and Adam Babiński and Antonio Polimeni and Maciej R Molas},
title = {Excitons and trions in WSSe monolayers},
journal = {2D Materials}
}

@Article{Lu_2020,
doi = {10.1088/2053-1583/ab5614},
year = {2019},
month = {nov},
publisher = {IOP Publishing},
volume = {7},
number = {1},
pages = {015017},
author = {Zhengguang Lu and Daniel Rhodes and Zhipeng Li and Dinh Van Tuan and Yuxuan Jiang and Jonathan Ludwig and Zhigang Jiang and Zhen Lian and Su-Fei Shi and James Hone and Hanan Dery and Dmitry Smirnov},
title = {Magnetic field mixing and splitting of bright and dark excitons in monolayer MoSe$_2$},
journal = {2D Materials}
}

@article{Kipczak2024,
	title   = {Impact of temperature on the brightening of neutral and charged dark excitons in {WSe$_2$} monolayer},
	author  = {Kipczak, {{\L}ucja} and Zawadzka, Natalia and Jana, Dipankar and Antoniazzi, Igor and Grzeszczyk, Magdalena and Zinkiewicz, Ma{\l}gorzata and Watanabe, Kenji and Taniguchi, Takashi and Potemski, Marek and Faugeras, Cl{\'e}ment and Babi{\'n}ski, Adam and Molas, Maciej R.},
	journal = {Nanophotonics},
	volume  = {13},
	number  = {26},
	pages   = {4743--4749},
	year    = {2024},
	doi     = {10.1515/nanoph-2024-0385}
}

@article{Robert2021,
author={Robert, Cedric
and Park, Sangjun
and Cadiz, Fabian
and Lombez, Laurent
and Ren, Lei
and Tornatzky, Hans
and Rowe, Alistair
and Paget, Daniel
and Sirotti, Fausto
and Yang, Min
and Van Tuan, Dinh
and Taniguchi, Takashi
and Urbaszek, Bernhard
and Watanabe, Kenji
and Amand, Thierry
and Dery, Hanan
and Marie, Xavier},
title={Spin/valley pumping of resident electrons in WSe$_2$ and WS$_2$ monolayers},
journal={Nature Communications},
year={2021},
month={Sep},
day={15},
volume={12},
number={1},
pages={5455},
issn={2041-1723},
doi={10.1038/s41467-021-25747-5}
}

@article{Arora2020,
  title = {Dark trions govern the temperature-dependent optical absorption and emission of doped atomically thin semiconductors},
  author = {Arora, Ashish and Wessling, Nils Kolja and Deilmann, Thorsten and Reichenauer, Till and Steeger, Paul and Kossacki, Piotr and Potemski, Marek and Michaelis de Vasconcellos, Steffen and Rohlfing, Michael and Bratschitsch, Rudolf},
  journal = {Phys. Rev. B},
  volume = {101},
  issue = {24},
  pages = {241413},
  numpages = {6},
  year = {2020},
  month = {Jun},
  publisher = {American Physical Society},
  doi = {10.1103/PhysRevB.101.241413}
}

@article{Slobodeniuk2016,
doi = {10.1088/2053-1583/3/3/035009},
year = {2016},
month = {aug},
publisher = {IOP Publishing},
volume = {3},
number = {3},
pages = {035009},
author = {Slobodeniuk, A O and Basko, D M},
title = {Spin–flip processes and radiative decay of dark intravalley excitons in transition metal dichalcogenide monolayers},
journal = {2D Materials}
}

@article{Robert2020,
author={Robert, C.
and Han, B.
and Kapuscinski, P.
and Delhomme, A.
and Faugeras, C.
and Amand, T.
and Molas, M. R.
and Bartos, M.
and Watanabe, K.
and Taniguchi, T.
and Urbaszek, B.
and Potemski, M.
and Marie, X.},
title={Measurement of the spin-forbidden dark excitons in MoS$_2$ and MoSe$_2$ monolayers},
journal={Nature Communications},
year={2020},
month={Aug},
day={12},
volume={11},
number={1},
pages={4037},
issn={2041-1723},
doi={10.1038/s41467-020-17608-4},
url={https://doi.org/10.1038/s41467-020-17608-4}
}

@Article{Kormanyos2015,
  author    = {Andor Korm{\'{a}}nyos and Guido Burkard and Martin Gmitra and Jaroslav Fabian and Viktor Z{\'{o}}lyomi and Neil D Drummond and Vladimir Fal'ko},
  journal   = {2D Materials},
  title     = {k $\cdotp$ p theory for two-dimensional transition metal dichalcogenide semiconductors},
  year      = {2015},
  month     = {apr},
  number    = {2},
  pages     = {022001},
  volume    = {2},
  doi       = {10.1088/2053-1583/2/2/022001},
  publisher = {{IOP} Publishing},
}

@Article{Li2018,
  author    = {Xiaoli Li and Yafang Shi and Shuai Li and Wei Shi and Wenpeng Han and Chuan Zhou and Xiaohui Zhao and Baolai Liang},
  journal   = {Optical Materials Express},
  title     = {Layer-number dependent reflection spectra of {MoS}$_2$ flakes on {SiO}$_2$/{Si} substrate},
  year      = {2018},
  month     = {sep},
  number    = {10},
  pages     = {3082},
  volume    = {8},
  doi       = {10.1364/ome.8.003082},
  publisher = {The Optical Society},
}

@article{Tongay2014,
    author = {Tongay, Sefaattin and Narang, Deepa S. and Kang, Jun and Fan, Wen and Ko, Changhyun and Luce, Alexander V. and Wang, Kevin X. and Suh, Joonki and Patel, K. D. and Pathak, V. M. and Li, Jingbo and Wu, Junqiao},
    title = {Two-dimensional semiconductor alloys: Monolayer Mo$_{1-x}$W$_x$Se$_2$},
    journal = {Applied Physics Letters},
    volume = {104},
    number = {1},
    pages = {012101},
    year = {2014},
    month = {01},
    issn = {0003-6951},
    doi = {10.1063/1.4834358},
    url = {https://doi.org/10.1063/1.4834358}
}

@article{Liu2013,
author={Liu, Wei
and Kang, Jiahao
and Sarkar, Deblina
and Khatami, Yasin
and Jena, Debdeep
and Banerjee, Kaustav},
title={Role of Metal Contacts in Designing High-Performance Monolayer n-Type WSe$_2$ Field Effect Transistors},
journal={Nano Letters},
year={2013},
month={May},
day={08},
publisher={American Chemical Society},
volume={13},
number={5},
pages={1983-1990},
issn={1530-6984},
doi={10.1021/nl304777e},
url={https://doi.org/10.1021/nl304777e}
}

@article{Echeverry2016,
  title = {Splitting between bright and dark excitons in transition metal dichalcogenide monolayers},
  author = {Echeverry, J. P. and Urbaszek, B. and Amand, T. and Marie, X. and Gerber, I. C.},
  journal = {Phys. Rev. B},
  volume = {93},
  issue = {12},
  pages = {121107},
  numpages = {5},
  year = {2016},
  month = {Mar},
  publisher = {American Physical Society},
  doi = {10.1103/PhysRevB.93.121107},
  url = {https://link.aps.org/doi/10.1103/PhysRevB.93.121107}
}

@article{Xiao2012,
  title = {Coupled Spin and Valley Physics in Monolayers of MoS$_2$ and Other Group-VI Dichalcogenides},
  author = {Xiao, Di and Liu, Gui-Bin and Feng, Wanxiang and Xu, Xiaodong and Yao, Wang},
  journal = {Phys. Rev. Lett.},
  volume = {108},
  issue = {19},
  pages = {196802},
  numpages = {5},
  year = {2012},
  month = {May},
  publisher = {American Physical Society},
  doi = {10.1103/PhysRevLett.108.196802},
  url = {https://link.aps.org/doi/10.1103/PhysRevLett.108.196802}
}

@article{Zhou2017,
author={Zhou, You
and Scuri, Giovanni
and Wild, Dominik S.
and High, Alexander A.
and Dibos, Alan
and Jauregui, Luis A.
and Shu, Chi
and De Greve, Kristiaan
and Pistunova, Kateryna
and Joe, Andrew Y.
and Taniguchi, Takashi
and Watanabe, Kenji
and Kim, Philip
and Lukin, Mikhail D.
and Park, Hongkun},
title={Probing dark excitons in atomically thin semiconductors via near-field coupling to surface plasmon polaritons},
journal={Nature Nanotechnology},
year={2017},
month={Sep},
day={01},
volume={12},
number={9},
pages={856-860},
issn={1748-3395},
doi={10.1038/nnano.2017.106},
url={https://doi.org/10.1038/nnano.2017.106}
}

@article{Kosmider2013,
  title = {Large spin splitting in the conduction band of transition metal dichalcogenide monolayers},
  author = {Ko\ifmmode \acute{s}\else \'{s}\fi{}mider, K. and Gonz\'alez, J. W. and Fern\'andez-Rossier, J.},
  journal = {Phys. Rev. B},
  volume = {88},
  issue = {24},
  pages = {245436},
  numpages = {7},
  year = {2013},
  month = {Dec},
  publisher = {American Physical Society},
  doi = {10.1103/PhysRevB.88.245436},
  url = {https://link.aps.org/doi/10.1103/PhysRevB.88.245436}
}

@article{Kopaczek2021,
author={Kopaczek, Jan
and Wo{\'{z}}niak, Tomasz
and Tamulewicz-Szwajkowska, Magdalena
and Zelewski, Szymon J.
and Serafi{\'{n}}czuk, Jaros{\l}aw
and Scharoch, Pawe{\l}
and Kudrawiec, Robert},
title={Experimental and Theoretical Studies of the Electronic Band Structure of Bulk and Atomically Thin Mo$_{1-x}$W$_x$Se$_2$ Alloys},
journal={ACS Omega},
year={2021},
month={Aug},
day={03},
publisher={American Chemical Society},
volume={6},
number={30},
pages={19893-19900},
doi={10.1021/acsomega.1c02788},
url={https://doi.org/10.1021/acsomega.1c02788}
}

@article{Slobodeniuk2019_BDXT,
doi = {10.1088/2053-1583/aaf9bb},
url = {https://doi.org/10.1088/2053-1583/aaf9bb},
year = {2019},
month = {jan},
publisher = {IOP Publishing},
volume = {6},
number = {2},
pages = {029501},
author = {Slobodeniuk, A O and Basko, D M},
title = {Corrigendum: Spin-flip processes and radiative decay of dark intravalley excitons in transition metal dichalcogenide monolayers (2016 2D Mater. 3 035009)},
journal = {2D Materials}
}

@article{Robert2017_DX,
  title = {Fine structure and lifetime of dark excitons in transition metal dichalcogenide monolayers},
  author = {Robert, C. and Amand, T. and Cadiz, F. and Lagarde, D. and Courtade, E. and Manca, M. and Taniguchi, T. and Watanabe, K. and Urbaszek, B. and Marie, X.},
  journal = {Phys. Rev. B},
  volume = {96},
  issue = {15},
  pages = {155423},
  numpages = {8},
  year = {2017},
  month = {Oct},
  publisher = {American Physical Society},
  doi = {10.1103/PhysRevB.96.155423},
  url = {https://link.aps.org/doi/10.1103/PhysRevB.96.155423}
}

@article{Molas_2017_DX_BDX,
doi = {10.1088/2053-1583/aa5521},
url = {https://doi.org/10.1088/2053-1583/aa5521},
year = {2017},
month = {jan},
publisher = {IOP Publishing},
volume = {4},
number = {2},
pages = {021003},
author = {Molas, M R and Faugeras, C and Slobodeniuk, A O and Nogajewski, K and Bartos, M and Basko, D M and Potemski, M},
title = {Brightening of dark excitons in monolayers of semiconducting transition metal dichalcogenides},
journal = {2D Materials}
}

@article{Schindler2026,
	author  = {Schindler, Janina J. and Jadczak, Joanna and Senki{\'c}, Ana and Bryja, Leszek and Wurstbauer, Ursula and A{\ss}mann, Marc and Debus, J{\"o}rg},
	title   = {Anomalous Valley Zeeman Splitting Induced by In-plane Magnetic Fields in a {Mo$_{0.7}$W$_{0.3}$Se$_2$} Monolayer},
	journal = {Advanced Quantum Technologies},
	volume  = {9},
	number  = {4},
	pages   = {e01019},
	year    = {2026},
	doi     = {10.1002/qute.202501019}
}

@book{Bir1974,
  title={Symmetry and Strain-induced Effects in Semiconductors},
  author={Bir, G.L. and Pikus, G.E.},
  isbn={9780470073216},
  lccn={74014842},
  series={A Halsted Press book},
  url={https://books.google.cz/books?id=38m2QgAACAAJ},
  year={1974},
  publisher={Wiley}
}

@article{OlkowskaPucko2026,
  title = {Extremely High Excitonic $g$ Factors in 2D Crystals by Alloy-Induced Admixing of Band States},
  author = {Olkowska-Pucko, Katarzyna and Wo\ifmmode \acute{z}\else \'{z}\fi{}niak, Tomasz and Blundo, Elena and Zawadzka, Natalia and Kipczak, \L{}ucja and Faria Junior, Paulo E. and Szpakowski, Jan and Krasucki, Grzegorz and Cianci, Salvatore and Vaclavkova, Diana and Jana, Dipankar and Kapu\ifmmode \acute{s}\else \'{s}\fi{}ci\ifmmode \acute{n}\else \'{n}\fi{}ski, Piotr and Pawbake, Amit and Badola, Shalini and Grzeszczyk, Magdalena and Cecchetti, Daniele and Pettinari, Giorgio and Antoniazzi, Igor and Sofer, Zden\ifmmode \check{e}\else \v{e}\fi{}k and Plutnarov\'a, Iva and Watanabe, Kenji and Taniguchi, Takashi and Faugeras, Clement and Potemski, Marek and Babi\ifmmode \acute{n}\else \'{n}\fi{}ski, Adam and Polimeni, Antonio and Molas, Maciej R.},
  journal = {Phys. Rev. Lett.},
  volume = {136},
  issue = {7},
  pages = {076901},
  numpages = {9},
  year = {2026},
  month = {Feb},
  publisher = {American Physical Society},
  doi = {10.1103/lx4n-7bb7},
  url = {https://link.aps.org/doi/10.1103/lx4n-7bb7}
}

@article{Popescu2012,
  title = {Extracting E versus stackrel P vec k effective band structure from supercell calculations on alloys and impurities},
  author = {Popescu, Voicu and Zunger, Alex},
  journal = {Phys. Rev. B},
  volume = {85},
  issue = {8},
  pages = {085201},
  numpages = {12},
  year = {2012},
  month = {Feb},
  publisher = {American Physical Society},
  doi = {10.1103/PhysRevB.85.085201},
  url = {https://link.aps.org/doi/10.1103/PhysRevB.85.085201}
}

@article{Kurdyubov2021,
  author = {Kurdyubov, A. S. and Trifonov, A. V. and Gerlovin, I. Ya. and
            Gribakin, B. F. and Grigoryev, P. S. and Mikhailov, A. V. and
            Ignatiev, I. V. and Efimov, Yu. P. and Eliseev, S. A. and
            Lovtcius, V. A. and A{\ss}mann, M. and Bayer, M. and Kavokin, A. V.},
  title = {Optical control of a dark exciton reservoir},
  journal = {Phys. Rev. B},
  volume = {104},
  pages = {035414},
  year = {2021},
  doi = {10.1103/PhysRevB.104.035414}
}

@article{Bayer2000,
  author = {Bayer, M. and Stern, O. and Kuther, A. and Forchel, A.},
  title = {Spectroscopic study of dark excitons in self-assembled quantum dots by a magnetic-field-induced symmetry breaking},
  journal = {Phys. Rev. B},
  volume = {61},
  pages = {7273},
  year = {2000},
  doi = {10.1103/PhysRevB.61.7273}
}

@article{Goryca2010,
  author = {Goryca, M. and Plochocka, P. and Kazimierczuk, T. and
            Wojnar, P. and Karczewski, G. and Gaj, J. A. and
            Potemski, M. and Kossacki, P.},
  title = {Brightening of dark excitons in a single {CdTe} quantum dot containing a single {Mn}$^{2+}$ ion},
  journal = {Phys. Rev. B},
  volume = {82},
  pages = {165323},
  year = {2010},
  doi = {10.1103/PhysRevB.82.165323}
}

@article{Huo2017,
  author = {Huo, Y. H. and Kr{\'a}pek, V. and Schmidt, O. G. and Rastelli, A.},
  title = {Spontaneous brightening of dark excitons in {GaAs/AlGaAs} quantum dots near a cleaved facet},
  journal = {Phys. Rev. B},
  volume = {95},
  pages = {165304},
  year = {2017},
  doi = {10.1103/PhysRevB.95.165304}
}

@article{Germanis2018,
  author = {Germanis, S. and Atkinson, P. and Hostein, R. and Gourdon, C. and
            Voliotis, V. and Lema{\^i}tre, A. and Bernard, M. and
            Margaillan, F. and Majrab, S. and Eble, B.},
  title = {Dark-bright exciton coupling in asymmetric quantum dots},
  journal = {Phys. Rev. B},
  volume = {98},
  pages = {155303},
  year = {2018},
  doi = {10.1103/PhysRevB.98.155303}
}

@article{Chen2024InSe,
  author = {Paylaga, Naomi Tabudlong and Chou, Chang-Ti and Lin, Chia-Chun and
            Taniguchi, Takashi and Watanabe, Kenji and Sankar, Raman and
            Chan, Yang-hao and Chen, Shao-Yu and Wang, Wei-Hua},
  title = {Monolayer indium selenide: an indirect bandgap material exhibits efficient brightening of dark excitons},
  journal = {npj 2D Materials and Applications},
  volume = {8},
  pages = {12},
  year = {2024},
  doi = {10.1038/s41699-024-00450-3}
}

@article{Sears2025CrSBr,
  author = {Sears, J. and Zager, B. and He, W. and Occhialini, C. A. and
            Shen, Y. and Lajer, M. and Villanova, J. W. and Berlijn, T. and
            Yakhou-Harris, F. and Brookes, N. B. and Chica, D. G. and
            Roy, X. and Baldini, E. and Pelliciari, J. and Bisogni, V. and
            Johnston, S. and Mitrano, M. and Dean, M. P. M.},
  title = {Observation of Anisotropic Dispersive Dark-Exciton Dynamics in {CrSBr}},
  journal = {Phys. Rev. Lett.},
  volume = {135},
  pages = {146503},
  year = {2025},
  doi = {10.1103/fz3h-6jdx}
}

@article{Kipczak2025,
  author = {Kipczak, {\L}ucja and Chen, Zhaolong and Grzeszczyk, Magdalena and
            Grebenchuk, Sergey and Huang, Pengru and Vaklinova, Kristina and
            Watanabe, Kenji and Taniguchi, Takashi and Babi{\'n}ski, Adam and
            Koperski, Maciej and Molas, Maciej R.},
  title = {Interplay between charge transfer and magnetic proximity effects
           in {WSe$_2$/CrCl$_3$} heterostructures},
  journal = {Nanoscale Horizons},
  volume = {10},
  number = {10},
  pages = {2465--2474},
  year = {2025},
  doi = {10.1039/D5NH00198F}
}

@article{FerroelectricBrightening2026,
  author = {Wang, Xinyun and Grzeszczyk, Magdalena and Trushin, Maxim and
            Verzhbitskiy, Ivan and Litvinov, Dmitrii and Ho, Yi Wei and
            Chen, Yuan and Wu, Zhenyue and Telychko, Mykola and
            Zhang, Chuanqi and Granados del Aguila, Andres and
            Goh, Kuan Eng Johnson and Li, Xinwei and Eda, Goki and
            Adam, Shaffique and Koperski, Maciej and Loh, Kian Ping},
  title = {Ferroelectric brightening of spin-forbidden dark excitons in a
           {WSe$_2$}/hybrid-perovskite heterostructure},
  journal = {Nature Communications},
  volume = {17},
  pages = {6285},
  year = {2026},
  doi = {10.1038/s41467-026-72143-y}
}

@article{PlasmonBrightening2025,
  author = {Jeong, Hyun and Suh, Hyeong Chan and Cho, Ga Hyun and
            Joo, Huitae and Koo, Yeonjeong and Ko, Hayoung and
            Kim, Ki Kang and Kim, Youngbum and Kim, Jeongyong and
            Park, Kyoung-Duck and Jeong, Mun Seok},
  title = {Large-Area Bright Emission of Plasmon-Coupled Dark Excitons
           at Room Temperature},
  journal = {Advanced Science},
  volume = {12},
  number = {3},
  pages = {2411841},
  year = {2025},
  doi = {10.1002/advs.202411841}
}

@article{Wiesner2022,
  author = {Wiesner, Maciej and Roberts, Richard H. and Ge, Ruijing and
            Mennel, Lukas and M{\"u}ller, Thomas and Lin, Jung-Fu and
            Akinwande, Deji and Jenczyk, Jacek},
  title = {Signatures of bright-to-dark exciton conversion in corrugated {MoS$_2$} monolayers},
  journal = {Applied Surface Science},
  volume = {600},
  pages = {154078},
  year = {2022},
  doi = {10.1016/j.apsusc.2022.154078}
}

@article{Janus2025,
  author = {Kopaczek, Jan and Hays, Patrick and Wu, Huan and Povilus, Blake and
            Erdi, Melike and Banerjee, Rounak and Wu, Cheng-Lun and
            Sanchez Esqueda, Ivan and Kudrawiec, Robert and
            Botana, Antia S. and Tongay, Seth Ariel},
  title = {Dark-to-Bright Exciton Transition in 2D Janus Excitonic Semiconductors via Metal Cation Alloying},
  journal = {ACS Nano},
  volume = {19},
  number = {44},
  pages = {38685--38695},
  year = {2025},
  doi = {10.1021/acsnano.5c13372}
}

@misc{krasucki2026dopinginducedbrighteningdarkexcitons,
      title={Doping-Induced Brightening of Dark Excitons and Trions in a WSe$_2$ Monolayer}, 
      author={Grzegorz Krasucki and Artur O. Slobodeniuk and Kacper Walczyk and Katarzyna Olkowska-Pucko and Kenji Watanabe and Takashi Taniguchi and Adam Babiński and Maciej R. Molas},
      year={2026},
      eprint={2604.25553},
      archivePrefix={arXiv},
      primaryClass={cond-mat.mes-hall},
      url={https://arxiv.org/abs/2604.25553}, 
}

@article{Dyksik2021,
  author  = {Dyksik, Mateusz and Duim, Herman and Maude, Duncan K. and Baranowski, Micha{\l} and Loi, Maria Antonietta and P{\l}ochocka, Paulina},
  title   = {Brightening of dark excitons in 2D perovskites},
  journal = {Science Advances},
  volume  = {7},
  number  = {46},
  pages   = {eabk0904},
  year    = {2021},
  doi     = {10.1126/sciadv.abk0904}
}

@article{Posmyk2022,
  author  = {Posmyk, Katarzyna and Zawadzka, Natalia and Dyksik, Mateusz and Surrente, Alessandro and Maude, Duncan K. and Kazimierczuk, Tomasz and Babi{\'n}ski, Adam and Molas, Maciej R. and Paritmongkol, Watcharaphol and M{\k{a}}czka, Miros{\l}aw and Tisdale, William A. and P{\l}ochocka, Paulina and Baranowski, Micha{\l}},
  title   = {Quantification of Exciton Fine Structure Splitting in a Two-Dimensional Perovskite Compound},
  journal = {The Journal of Physical Chemistry Letters},
  volume  = {13},
  number  = {20},
  pages   = {4463--4469},
  year    = {2022},
  doi     = {10.1021/acs.jpclett.2c00942}
}

@article{Thompson2024,
  author  = {Thompson, Joshua J. P. and Dyksik, Mateusz and Peksa, Paulina and Posmyk, Katarzyna and Joki, Ambj{\"o}rn and Perea-Causin, Raul and Erhart, Paul and Baranowski, Micha{\l} and Loi, Maria Antonietta and P{\l}ochocka, Paulina and Malic, Ermin},
  title   = {Phonon-Bottleneck Enhanced Exciton Emission in 2D Perovskites},
  journal = {Advanced Energy Materials},
  volume  = {14},
  number  = {20},
  pages   = {2304343},
  year    = {2024},
  doi     = {10.1002/aenm.202304343}
}

@misc{SM,
  note = {See Supplemental Material for details of sample fabrication and magneto-optical measurements, band-structure calculations, the derivation of the collected-emission intensity, further theoretical discussion, and the sensitivity of the extracted in-plane $g$ factor to the bright-dark exciton splitting.}
}

@article{Giannozzi2009,
  author  = {Giannozzi, Paolo and others},
  title   = {{QUANTUM ESPRESSO}: a modular and open-source software project
             for quantum simulations of materials},
  journal = {J. Phys.: Condens. Matter},
  volume  = {21},
  pages   = {395502},
  year    = {2009},
  doi     = {10.1088/0953-8984/21/39/395502}
}

@article{Giannozzi2017,
  author  = {Giannozzi, Paolo and others},
  title   = {Advanced capabilities for materials modelling with
             {Quantum ESPRESSO}},
  journal = {J. Phys.: Condens. Matter},
  volume  = {29},
  pages   = {465901},
  year    = {2017},
  doi     = {10.1088/1361-648X/aa8f79}
}

@article{Giannozzi2020,
  author  = {Giannozzi, Paolo and others},
  title   = {{Quantum ESPRESSO} toward the exascale},
  journal = {J. Chem. Phys.},
  volume  = {152},
  pages   = {154105},
  year    = {2020},
  doi     = {10.1063/5.0005082}
}

@article{Medeiros2015,
  author  = {Medeiros, Paulo V. C. and Tsirkin, Stepan S. and
             Stafstr{\"o}m, Sven and Bj{\"o}rk, Jonas},
  title   = {Unfolding spinor wave functions and expectation values
             of general operators: Introducing the unfolding-density operator},
  journal = {Phys. Rev. B},
  volume  = {91},
  pages   = {041116(R)},
  year    = {2015},
  doi     = {10.1103/PhysRevB.91.041116}
}

@article{Marini2009,
  author  = {Marini, Andrea and Hogan, Conor and Gr{\"u}ning, Myrta
             and Varsano, Daniele},
  title   = {{Yambo}: An ab initio tool for excited state calculations},
  journal = {Comput. Phys. Commun.},
  volume  = {180},
  pages   = {1392--1403},
  year    = {2009},
  doi     = {10.1016/j.cpc.2009.02.003}
}

@article{Sangalli2019,
  author  = {Sangalli, Davide and others},
  title   = {Many-body perturbation theory calculations using the
             {yambo} code},
  journal = {J. Phys.: Condens. Matter},
  volume  = {31},
  pages   = {325902},
  year    = {2019},
  doi     = {10.1088/1361-648X/ab15d0}
}

@article{PhysRevLett.77.3865,
  title = {Generalized Gradient Approximation Made Simple},
  author = {Perdew, John P. and Burke, Kieron and Ernzerhof, Matthias},
  journal = {Phys. Rev. Lett.},
  volume = {77},
  issue = {18},
  pages = {3865--3868},
  numpages = {0},
  year = {1996},
  month = {Oct},
  publisher = {American Physical Society},
  doi = {10.1103/PhysRevLett.77.3865},
  url = {https://link.aps.org/doi/10.1103/PhysRevLett.77.3865}
}

@article{PhysRevB.88.085117,
  title = {Optimized norm-conserving Vanderbilt pseudopotentials},
  author = {Hamann, D. R.},
  journal = {Phys. Rev. B},
  volume = {88},
  issue = {8},
  pages = {085117},
  numpages = {10},
  year = {2013},
  month = {Aug},
  publisher = {American Physical Society},
  doi = {10.1103/PhysRevB.88.085117},
  url = {https://link.aps.org/doi/10.1103/PhysRevB.88.085117}
}

@article{PhysRevB.78.085125,
  title = {Accurate $GW$ self-energies in a plane-wave basis using only a few empty states: Towards large systems},
  author = {Bruneval, Fabien and Gonze, Xavier},
  journal = {Phys. Rev. B},
  volume = {78},
  issue = {8},
  pages = {085125},
  numpages = {9},
  year = {2008},
  month = {Aug},
  publisher = {American Physical Society},
  doi = {10.1103/PhysRevB.78.085125},
  url = {https://link.aps.org/doi/10.1103/PhysRevB.78.085125}
}

@article{Cianci2023,
  author = {Cianci, Salvatore and Blundo, Elena and Tuzi, Federico and Pettinari, Giorgio and Olkowska-Pucko, Katarzyna and Parmenopoulou, Eirini and Peeters, Djero B. L. and Miriametro, Antonio and Taniguchi, Takashi and Watanabe, Kenji and Babinski, Adam and Molas, Maciej R. and Felici, Marco and Polimeni, Antonio},
  title = {Spatially Controlled Single Photon Emitters in {hBN}-Capped {WS2} Domes},
  journal = {Advanced Optical Materials},
  volume = {11},
  number = {12},
  pages = {2202953},
  year = {2023},
  doi = {10.1002/adom.202202953}
}

\newpage
\onecolumngrid

\renewcommand{\thefigure}{S\arabic{section}.\arabic{figure}}
\renewcommand{\thesection}{S\arabic{section}}

\begin{center}
	{\large{ {\bf Supplemental Material:\\  Magnetic-Field-Induced Darkening of Bright Excitons \\ in a MoWSe$_2$ Monolayer Alloy}}}
	\vskip0.5\baselineskip{Katarzyna Olkowska-Pucko,{$^{1}$} Artur O. Slobodeniuk,{$^{2,3}$} Mehdi Arfaoui,{$^{4,5}$} Elena Blundo,{$^{6}$}
		Natalia Zawadzka,{$^{1}$} Łucja Kipczak,{$^{1}$} Jan Szpakowski,{$^{1}$} Diana Vaclavkova,{$^{7,8}$} Dipankar Jana,{$^{7,8}$}
		Piotr Kapuściński,{$^{1,7}$} Kenji Watanabe,{$^{9}$} Takashi Taniguchi,{$^{10}$} Clement Faugeras,{$^{7}$}
		Marek Potemski,{$^{1,7,3}$} Adam Babiński,{$^{1}$} Antonio Polimeni,{$^{6}$} and Maciej R. Molas{$^{1}$}}
	
	\vskip0.5\baselineskip{\itshape$^{1}$ University of Warsaw, Faculty of Physics, 02-093 Warsaw, Poland \\$^{2}$ Charles University, Faculty of Mathematics and Physics,  CZ-121 16 Prague, Czech Republic 
		\\$^{3}$ CENTERA, CEZAMAT, Warsaw University of Technology, 02-822 Warsaw, Poland
		\\$^{4}$ Université Paris-Saclay, CEA, CNRS, SPEC, 91191 Gif-sur-Yvette, France
		\\$^{5}$ Université Paris-Saclay, ONERA, CNRS, Laboratoire d'étude des microstructures (LEM), 92322 Châtillon, France
		\\$^{6}$ Physics Department, Sapienza University of Rome, 00185 Rome, Italy
		\\$^{7}$ Laboratoire National des Champs Magnétiques Intenses, CNRS-UGA-UPS-INSA-EMFL, 38042 Grenoble, France
		\\$^{8}$ Institute for Functional Intelligent Materials, National University of Singapore, 117544, Singapore
		\\$^{9}$ Research Center for Electronic and Optical Materials, National Institute for Materials Science, 1-1 Namiki, Tsukuba 305-0044, Japan 
		\\$^{10}$ Research Center for Materials Nanoarchitectonics, National Institute for Materials Science, 1-1 Namiki, Tsukuba 305-0044, Japan}
\end{center}

This Supplemental Material provides: \ref{Methods} - Fabrication of the hBN-encapsulated Mo$_{0.68}$W$_{0.32}$Se$_2$ monolayer, the magneto-photoluminescence measurements performed in high in-plane magnetic fields, and computational details. \ref{sec:SM_theory} - Emission intensity of field-mixed bright and dark excitons. \ref{sec:SM_PL} - Logarithmic-scale representation of the magneto-photoluminescence spectra highlighting the weak X$^\mathrm{D}$ emission. \ref{sec:SM_band} - Band structure and spin-resolved unfolding of monolayer Mo$_{0.75}$W$_{0.25}$Se$_2$. \ref{sec:SM_discussion} - Effective-medium description of magnetic darkening in Mo$_{0.68}$W$_{0.32}$Se$_2$. \ref{sec:SM_sensitivity} - Sensitivity of the extracted $g_\parallel$ factor to the bright-dark exciton splitting.

\renewcommand{\thesection}{Methods}
\section{\label{Methods}}

\noindent
\textbf{Fabrication of hBN encapsulated sample}\\
Mo$_{0.68}$W$_{0.32}$Se$_2$ was mechanically exfoliated from a bulk crystal purchased from HQ Graphene.
The monolayer (ML) was exfoliated onto polydimethylsiloxane (PDMS) using scotch tape.
Relatively thick hBN flakes were mechanically exfoliated onto SiO$_2$/Si substrates consisting of a $300~\mathrm{nm}$-thick SiO$_2$ layer on a $500~\mu\mathrm{m}$-thick Si wafer.
The Mo$_{0.68}$W$_{0.32}$Se$_2$ ML was then deterministically transferred from the PDMS onto an hBN flake.
A thin hBN flake was exfoliated onto PDMS and subsequently transferred on top of the ML to complete the encapsulation.
After each transfer step, the sample was annealed under high vacuum ($10^{-6}$~mbar) at approximately $120^\circ$C for several hours to promote the coalescence of contaminants and improve the adhesion between the layers.
\\

\noindent
\textbf{Magneto-optical experiments}\\
Low-temperature micro-magneto-photoluminescence investigations were performed in the Voigt geometry, $i.e.$, with the magnetic field oriented parallel to the ML plane.
Measurements with a spatial resolution of approximately $1~\mu$m were carried out using a resistive magnet capable of producing magnetic fields up to 30~T and a free-space optical setup.
The sample was mounted on an $x$-$y$-$z$ piezoelectric stage and maintained at $T=4.2$~K.
The sample was excited using a continuous-wave (CW) laser diode with a wavelength of 515~nm (photon energy of 2.41~eV).
The emitted light was dispersed by a monochromator with a focal length of 0.5~m and detected using a CCD camera.
\\

\noindent
\textbf{Computational details}\\
First-principles calculations were performed within (density-functional-theory) DFT using the plane-wave pseudopotential approach as implemented in the \textsc{Quantum ESPRESSO} package~\cite{Giannozzi2009,Giannozzi2017,Giannozzi2020}.
The exchange-correlation interaction was described within the generalized-gradient approximation using the Perdew-Burke-Ernzerhof functional~\cite{PhysRevLett.77.3865}. Optimized norm-conserving Vanderbilt pseudopotentials were employed for Mo, W, and Se.
The Mo$_{0.75}$W$_{0.25}$Se$_2$ alloy was represented by a $2\times2\times1$ supercell (SC) containing four transition-metal sites (three Mo and one W) and eight Se atoms~\cite{PhysRevB.88.085117}.
The alloy configuration was constructed by replacing one of the four
Mo atoms in the $2\times2$ MoSe$_2$ SC by W.
The atomic positions and in-plane lattice parameters were optimized
within DFT using Broyden-Fletcher-Goldfarb-Shanno algorithm, while the out-of-plane lattice parameter was kept fixed at 20~\AA{}.
Structural relaxation was performed until the total-energy and force convergence thresholds of $10^{-4}$~Ry and $10^{-3}$~Ry/Bohr, respectively. The optimized in-plane SC lattice constant was $6.64$~\AA{}.
The Brillouin zone (BZ) was sampled with a $12\times12\times1$ $\mathbf{k}$-point grid during
the structural relaxation.
For the subsequent electronic-structure calculations, the Kohn-Sham
wave functions were expanded in a plane-wave basis with a kinetic-energy
cutoff of $60$~Ry. The BZ was sampled using an $18\times18\times1$ $\mathbf{k}$-point grid. The self-consistent-field calculations were performed with a convergence threshold of $10^{-8}$~Ry. Spin-orbit coupling (SOC) was explicitly included in the electronic-structure calculations. A two-dimensional isolated-system treatment was employed along the out-of-plane direction.

Quasiparticle corrections were calculated within the $G_0W_0$ approximation using \textsc{Yambo}~\cite{Marini2009,Sangalli2019}. The dynamically screened Coulomb interaction was described within the plasmon-pole approximation. The independent-particle polarizability included $400$ bands, and the dielectric matrix was expanded using a cutoff of $10$~Ry.
The convergence of the self-energy with respect to high-energy
unoccupied states was accelerated using the Bruneval-Gonze terminator~\cite{PhysRevB.78.085125}.
A slab Coulomb cutoff was applied along the out-of-plane direction
to suppress spurious interactions between periodically repeated
MLs.

\noindent

\clearpage
\newpage
\renewcommand{\thesection}{S\arabic{section}}
\setcounter{section}{0}
\setcounter{figure}{0}

\clearpage
\newpage
\section{Emission intensity of field-mixed bright and dark excitons}
\label{sec:SM_theory}

In this section, we examine how the magnetic-field-induced mixing of bright and dark excitons modifies the transition dipole moment and, consequently, the light intensity collected in the experimental geometry.
In particular, we derive the dependence of the collected emission intensity on the mixing angle and show that the modification of the transition dipole alone produces only a weak reduction of the bright-exciton emission under our experimental conditions.

The field-mixed bright-like state introduced in the main text is characterized by the transition dipole moment
$\mathbf{d}_\pm=\langle 0|\widehat{\mathbf{d}}|\mathrm{B}_\pm\rangle$,
which has the form
\begin{align}
\mathbf{d}_\pm=&-i\cos\Big(\frac{\vartheta}{2}\Big)
d_\parallel(\pm\mathbf{e}_x+i\mathbf{e}_y)
+ie^{\pm i\phi}\sin\Big(\frac{\vartheta}{2}\Big)
d_\perp\mathbf{e}_z.
\end{align}
The dipole vectors associated with the two valleys can generally be written as
$\mathbf{d}_\alpha=a_\alpha[\mathbf{e}_x+i\mathbf{e}_y]
+b_\alpha[-\mathbf{e}_x+i\mathbf{e}_y]
+c_\alpha\mathbf{e}_z$, with $\alpha=\pm$.

We now derive the intensity of circularly polarized emission as a function of the collection angle and the orientation of the dipole moments.
The derivation is based on a transformation to the spherical basis and takes into account the directional character of optical-dipole emission.
The resulting expressions provide a quantitative description of the influence of the finite collection angle of the detection system on the measured emission intensity.

\begin{figure}[h!]
    \centering
    \includegraphics[width=0.75\linewidth]{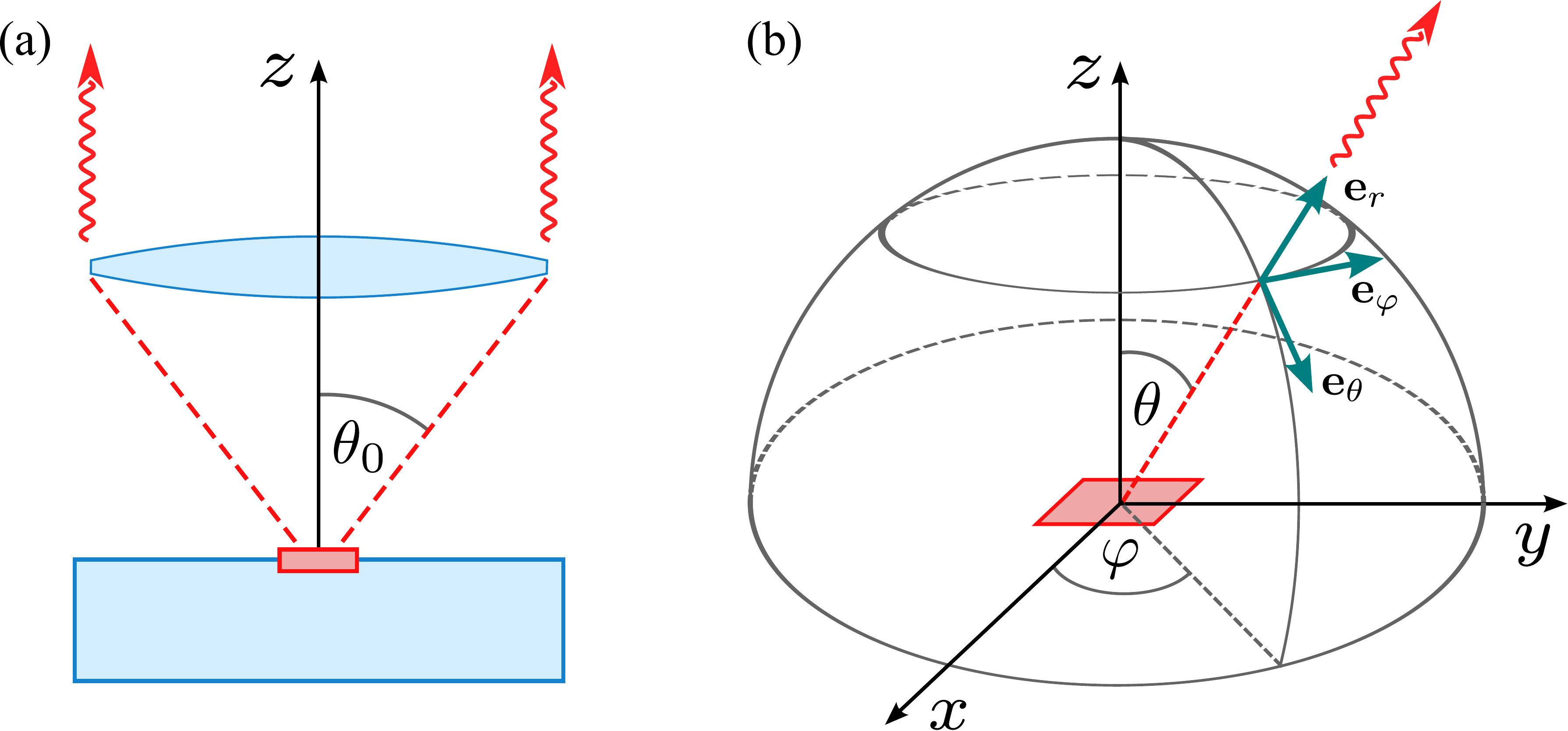}
    \caption{
    (a) Side view of the relevant part of the optical setup.
    The red dashed lines delimit the collection cone with half-angle $\theta_0$; photons emitted within this cone are collected by the lens.
    The wavy arrows represent the light emerging from the lens.
    The red rectangle at the focal plane represents the S-TMD sample placed on the substrate, shown as a blue rectangle.
    (b) Schematic representation of a photon emitted from the sample in the $(\theta,\varphi)$ direction.
    The three teal arrows indicate the local spherical basis vectors.
    }
    \label{fig:fig_1}
\end{figure}

Here $\mathbf{e}_\pm(\theta,\varphi)$ are the polarization vectors of $\sigma_\pm$-polarized photons propagating in the direction defined by the angles $(\theta,\varphi)$.
To write $\mathbf{e}_\pm$ explicitly, we introduce the local orthogonal unit vectors in the directions of increasing $r$, $\theta$, and $\varphi$, respectively [see Fig.~\ref{fig:fig_1}(b)]:
\begin{align}
&\mathbf{e}_r=\sin\theta\cos\varphi\,\mathbf{e}_x +\sin\theta\sin\varphi\,\mathbf{e}_y +\cos\theta\,\mathbf{e}_z,\\
&\mathbf{e}_\theta=\cos\theta\cos\varphi\,\mathbf{e}_x +\cos\theta\sin\varphi\,\mathbf{e}_y -\sin\theta\,\mathbf{e}_z,\\
&\mathbf{e}_\varphi=-\sin\varphi\,\mathbf{e}_x +\cos\varphi\,\mathbf{e}_y. 
\end{align}
The spherical basis $\{\mathbf{e}_\theta,\mathbf{e}_\varphi,\mathbf{e}_r\}$ has the same relative orientation as $\{\mathbf{e}_x,\mathbf{e}_y,\mathbf{e}_z\}$.
Hence, we define $\mathbf{e}_\pm(\theta,\varphi)\equiv (\pm\mathbf{e}_\theta+i\mathbf{e}_\varphi)/\sqrt{2}$.
Taking the corresponding scalar products gives
\begin{align}
\mathbf{d}_\alpha\cdot\mathbf{e}_\pm(\theta,\varphi) =&-\frac{a_\alpha}{\sqrt{2}}(1\mp\cos\theta)e^{i\varphi} -\frac{b_\alpha}{\sqrt{2}}(1\pm\cos\theta)e^{-i\varphi}
\nonumber\\
&\mp\frac{c_\alpha}{\sqrt{2}}\sin\theta.
\end{align}

For the two valleys, we obtain
\begin{align}
\mathbf{d}_+\cdot\mathbf{e}_\pm(\theta,\varphi) =&\frac{id_\parallel}{\sqrt{2}}(1\mp\cos\theta) \cos\Big(\frac{\vartheta}{2}\Big)e^{i\varphi}
\nonumber\\
&\mp\frac{id_\perp}{\sqrt{2}} \sin\theta\sin\Big(\frac{\vartheta}{2}\Big)e^{i\phi},
\end{align}
and
\begin{align}
\mathbf{d}_-\cdot\mathbf{e}_\pm(\theta,\varphi) =&\frac{id_\parallel}{\sqrt{2}}(1\pm\cos\theta) \cos\Big(\frac{\vartheta}{2}\Big)e^{-i\varphi}
\nonumber\\
&\mp\frac{id_\perp}{\sqrt{2}} \sin\theta\sin\Big(\frac{\vartheta}{2}\Big)e^{-i\phi}.
\end{align}

Here, $\phi$ denotes the azimuthal orientation of the in-plane magnetic field, whereas $\varphi$ denotes the azimuthal direction of photon emission.

Squaring these amplitudes gives

\begin{align}
|\mathbf{d}_+\cdot\mathbf{e}_\pm(\theta,\varphi)|^2 =&\frac{d^2_\parallel}{2}(1\mp\cos\theta)^2 \cos^2\Big(\frac{\vartheta}{2}\Big)
+\frac{d^2_\perp}{2}\sin^2\theta \sin^2\Big(\frac{\vartheta}{2}\Big)
\nonumber\\
&\mp d_\parallel d_\perp(1\mp\cos\theta)\sin\theta \sin\Big(\frac{\vartheta}{2}\Big) \cos\Big(\frac{\vartheta}{2}\Big)
\cos(\varphi-\phi),
\\
|\mathbf{d}_-\cdot\mathbf{e}_\pm(\theta,\varphi)|^2 =&\frac{d^2_\parallel}{2}(1\pm\cos\theta)^2 \cos^2\Big(\frac{\vartheta}{2}\Big)
+\frac{d^2_\perp}{2}\sin^2\theta \sin^2\Big(\frac{\vartheta}{2}\Big)
\nonumber\\
&\mp d_\parallel d_\perp(1\pm\cos\theta)\sin\theta \sin\Big(\frac{\vartheta}{2}\Big) \cos\Big(\frac{\vartheta}{2}\Big)
\cos(\varphi-\phi).
\end{align}

These expressions determine the intensity of the $\sigma_\pm$-polarized light emitted by excitons from the $\mathbf{K}_\pm$ valleys.
The total intensity of the $\sigma_\pm$-polarized light collected within the solid angle of the objective is therefore
\begin{equation}
I(\sigma_\pm) \propto \int_0^{2\pi}d\varphi \int_0^{\theta_0}\sin\theta\,d\theta \sum_{\alpha} |\mathbf{d}_\alpha\cdot\mathbf{e}_\pm(\theta,\varphi)|^2.
\end{equation}

Evaluating the sums over $\alpha$, we obtain

\begin{align}
\sum_\alpha|\mathbf{d}_\alpha\cdot\mathbf{e}_+(\theta,\varphi)|^2 =&d^2_\parallel(1+\cos^2\theta) \cos^2\Big(\frac{\vartheta}{2}\Big)
+d^2_\perp\sin^2\theta \sin^2\Big(\frac{\vartheta}{2}\Big)
\nonumber\\
&-2d_\parallel d_\perp\sin\theta \sin\Big(\frac{\vartheta}{2}\Big) \cos\Big(\frac{\vartheta}{2}\Big) \cos(\varphi-\phi),
\\
\sum_\alpha|\mathbf{d}_\alpha\cdot\mathbf{e}_-(\theta,\varphi)|^2 =&d^2_\parallel(1+\cos^2\theta) \cos^2\Big(\frac{\vartheta}{2}\Big)
+d^2_\perp\sin^2\theta \sin^2\Big(\frac{\vartheta}{2}\Big)
\nonumber\\
&+2d_\parallel d_\perp\sin\theta \sin\Big(\frac{\vartheta}{2}\Big) \cos\Big(\frac{\vartheta}{2}\Big) \cos(\varphi-\phi).
\end{align}

Integration over $\varphi$ eliminates the interference term in both expressions and demonstrates that the total $\sigma_+$ and $\sigma_-$ intensities are equal:
\begin{align}
I(\sigma_\pm) =&\mathcal{N}\int_0^{\theta_0}d\theta\,\sin\theta \Bigg[ d^2_\parallel(1+\cos^2\theta) \cos^2\Big(\frac{\vartheta}{2}\Big)
+d^2_\perp\sin^2\theta \sin^2\Big(\frac{\vartheta}{2}\Big) \Bigg]
\nonumber\\
=&\mathcal{N}\frac{d^2_\parallel}{3} \Big[4-3\cos(\theta_0)-\cos^3(\theta_0)\Big] \cos^2\Big(\frac{\vartheta}{2}\Big)
\nonumber\\
&+\mathcal{N}\frac{4d^2_\perp}{3} \sin^4\left(\frac{\theta_0}{2}\right) \Big[2+\cos(\theta_0)\Big] \sin^2\Big(\frac{\vartheta}{2}\Big).
\end{align}

Here, $\mathcal{N}$ is a coefficient independent of $\theta_0$ and $\vartheta$ and proportional to the bright-exciton population,
$\mathcal{N}\propto n_\mathrm{B}$.
In the limit $d_\parallel\gg d_\perp$, the collected emission intensity as a function of the in-plane magnetic field $B_\parallel$ is given by
\begin{align}
I(\sigma_\pm,B_\parallel) \propto \cos^2\Big(\frac{\vartheta}{2}\Big) = \frac{1}{2} +\frac{1}{2} \frac{\Delta} {\sqrt{\Delta^2+(g_\parallel\mu_BB_\parallel)^2}}.
\end{align}

The reduction in the collected emission intensity predicted by this expression originates from the field-induced rotation of the transition dipole moment of the bright-like exciton toward the out-of-plane direction.
However, even at high magnetic fields ($B_\parallel\sim30~\mathrm{T}$), this effect produces only a minor decrease in the collected A-exciton emission intensity:
\begin{align}
\frac{I(\sigma_\pm,30~\mathrm{T})} {I(\sigma_\pm,0)} \approx0.989.
\end{align}

\clearpage
\newpage

\section{Magnetic brightening of the dark excition}
\label{sec:SM_PL}

\begin{figure}[!h]
    \centering
    \includegraphics[width=0.6\linewidth]{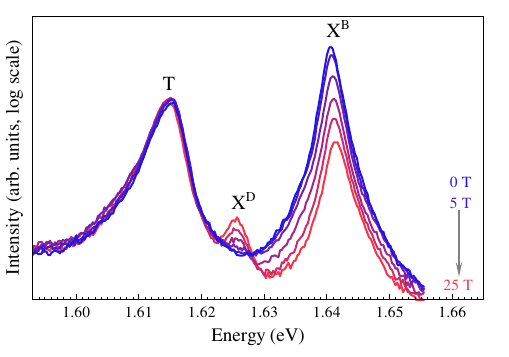}
    \caption{Evolution of the low-temperature photoluminescence spectra of the Mo$_{0.68}$W$_{0.32}$Se$_2$ ML measured in an in-plane magnetic field up to 25~T, shown on a logarithmic intensity scale to highlight the weak X$^\mathrm{D}$ emission. The excitation energy and laser power were 2.41~eV and 10~$\mu$W, respectively.}
    \label{fig:PL1}
\end{figure}

To highlight the weak X$^\mathrm{D}$ emission, Fig.~\ref{fig:PL1} presents the same PL spectra as in the main text on a logarithmic intensity scale.
This representation more clearly reveals the low-intensity X$^\mathrm{D}$ line emerging below X$^\mathrm{B}$ for $B_\parallel \gtrsim 15~\mathrm{T}$.

\clearpage
\newpage

\section{BAND STRUCTURE AND SPIN-RESOLVED UNFOLDING OF MONOLAYER Mo$_{0.75}$W$_{0.25}$Se$_2$}
\label{sec:SM_band}

The emergence of X$^{\mathrm{D}}$ below X$^{\mathrm{B}}$ in an in-plane magnetic field suggests that the lowest-energy excitonic state in the investigated Mo$_{0.68}$W$_{0.32}$Se$_2$ monolayer (ML) is optically forbidden. 
In particular, at $B_{\parallel}=25$ T, X$^{\mathrm{D}}$ lies approximately 15 meV below X$^{\mathrm{B}}$ (see Fig.~2 of the main text), while its
intensity increases quadratically with $B_{\parallel}$, as expected for the magnetic brightening of a spin-forbidden dark exciton. 
To further substantiate this assignment, we performed spin-resolved electronic-structure calculations to determine the band-edge spin ordering at the $\mathbf{K}^\pm$ valleys.
The calculations were performed for ML Mo$_{0.75}$W$_{0.25}$Se$_2$, represented by a $2\times2\times1$ alloy SC,  which provides a commensurate approximation to the experimentally investigated Mo$_{0.68}$W$_{0.32}$Se$_2$ composition.
Our calculations are intended to identify the relevant band-edge electronic structure rather than to reproduce quantitatively the experimentally determined bright-dark exciton splitting.\\

\begin{figure}[!h]
    \centering
    \includegraphics[width=0.4\linewidth]{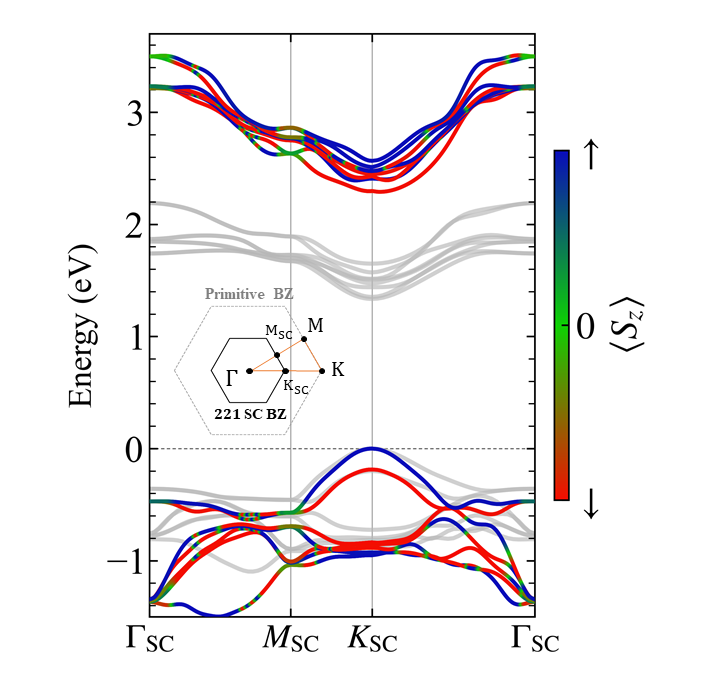}
    \caption{Spin-resolved $G_0W_0$ quasiparticle band structure of ML Mo$_{0.75}$W$_{0.25}$Se$_2$ calculated using a $2\times2\times1$ SC with SOC along the SC $\Gamma_{SC}$M$_{SC}$K$_{SC}$$\Gamma_{SC}$.
    The color scale represents the out-of-plane spin polarization, with red and blue corresponding to opposite spin projections and green to $\langle S_z \rangle\simeq0$. The energy zero is set to the valence-band maximum.
    The grey lines represent the DFT-calculated band structure.}
    \label{fig:GW_SOC}
\end{figure}

\noindent
\textbf{$\bm{G_0W_0}$ quasiparticle band structure and spin ordering}\\ 
The spin-resolved $G_0W_0$ quasiparticle band structure of ML Mo$_{0.75}$W$_{0.25}$Se$_2$ is shown in Fig.~\ref{fig:GW_SOC}.
Details of the computational method are provided in the Methods section.
Because the calculations are performed using a $2\times2\times1$ alloy SC, the quasiparticle bands are folded into the reduced SC Brillouin zone.
We denote the primitive-cell (PC) reciprocal-lattice basis vectors by $\mathbf{b}_i$ and the corresponding SC reciprocal-lattice basis vectors by $\mathbf{B}_i$, where $i=1,2,3$.
The real-space PC and SC lattice vectors are related by the transformation matrix $\mathbf{M}=\mathrm{diag}(2,2,1)$.
Accordingly, in reciprocal space, $\mathbf{B}_1=\mathbf{b}_1/2$, $\mathbf{B}_2=\mathbf{b}_2/2$, and $\mathbf{B}_3=\mathbf{b}_3$.
The quasiparticle band structure is evaluated along the SC path
$\Gamma_{\mathrm{SC}}M_{\mathrm{SC}}K_{\mathrm{SC}}\Gamma_{\mathrm{SC}}$,
where $\Gamma_{\mathrm{SC}}=(0,0,0)$, $M_{\mathrm{SC}}=(1/2,0,0)$, and $K_{\mathrm{SC}}=(1/3,2/3,0)$ are given in fractional coordinates of the SC reciprocal basis $\{\mathbf{B}_i\}$.
The inset of Fig.~\ref{fig:GW_SOC} compares the first SC BZ with the first PC BZ.
The DFT calculation including SOC yields a direct fundamental band gap at $K$ of $1.33$~eV.
The $G_0W_0$ correction increases the direct quasiparticle gap at $K$ to $2.29$~eV.

\noindent
\textbf{Spin-resolved band unfolding}\\
The SC folds states originating from different PC momenta into the reduced SC BZ, complicating the identification of their valley and spin character. 
To overcome this, we employ a spin-resolved band-unfolding procedure applied to fully relativistic DFT+SOC wave functions. 
The implementation follows the plane-wave spectral-weight formulation of SC band unfolding~\cite{Popescu2012}, generalized to two-component spinor wave functions as in Medeiros \textit{et al.}~\cite{Medeiros2015}.
For an arbitrary SC wave vector $\mathbf{K}$, the corresponding PC momenta satisfy $\mathbf{K}+\mathbf{G}=\mathbf{k}+\mathbf{g}$, where $\mathbf{G}=\sum_i g_i\mathbf{B}_i$ and $\mathbf{g}=\sum_i m_i\mathbf{b}_i$ are SC and PC reciprocal-lattice vectors, respectively, with $g_i,m_i\in \mathbb{Z}$.
Equivalently, a given SC momentum $\mathbf{K}$ unfolds onto the PC momenta $\mathbf{k}_j=\mathbf{K}+\mathbf{G}_j$ modulo a PC reciprocal-lattice vector, where $\mathbf G_j$ denotes one of the distinct SC reciprocal-lattice translations connecting $\mathbf K$ to the PC BZ.
A fully relativistic SC eigenstate is expressed as the two-component spinor
$\Psi_{n\mathbf{K}}(\mathbf{r})
=(\psi_{n\mathbf{K}}^{\uparrow}(\mathbf{r}),
\psi_{n\mathbf{K}}^{\downarrow}(\mathbf{r}))^{\mathrm{T}}$,
where $n$ denotes the SC band index. Each spinor component is expanded in a plane-wave basis as
$\psi_{n\mathbf{K}}^{\sigma}(\mathbf{r})
=\sum_{\mathbf{G}}C_{n\mathbf{K}}^{\sigma}(\mathbf{G})
e^{i(\mathbf{K}+\mathbf{G})\cdot\mathbf{r}}$,
with $\sigma=\uparrow,\downarrow$. The spin-resolved unfolded spectral weight at a given PC momentum $\mathbf{k}$ is obtained by summing the plane-wave contributions over the subset $\mathcal{U}_{\mathbf{k}}$ of SC reciprocal-lattice vectors satisfying the unfolding condition,
\begin{equation}
W_{n\mathbf{K}}^{\sigma}(\mathbf{k})
=
\sum_{\mathbf{G}\in\mathcal{U}_{\mathbf{k}}}
\left|C_{n\mathbf{K}}^{\sigma}(\mathbf{G})\right|^2,
\label{eq:spin_weight}
\end{equation}
The total unfolded spectral weight is
\begin{equation}
W_{n\mathbf{K}}(\mathbf{k})
=
W_{n\mathbf{K}}^{\uparrow}(\mathbf{k})
+
W_{n\mathbf{K}}^{\downarrow}(\mathbf{k}).
\label{eq:total_weight}
\end{equation}
For the present $2\times2\times1$ transformation, $\det\mathbf{M}=4$, giving four distinct unfolding sectors for each SC wave vector $\mathbf K$.
In our plane-wave implementation, the subset $\mathcal{U}_{\mathbf{k}}$ entering Eq.~\eqref{eq:spin_weight} is identified from the integer coefficients $(g_1,g_2,g_3)$ of the SC reciprocal-lattice vectors.
Because the SC is doubled along the first two lattice directions, the unfolding sectors are determined by the parities of $g_1$ and $g_2$. 
We therefore label each sector by $(s_1,s_2)\in\{(0,0),(1,0),(0,1),(1,1)\}$, such that the selected plane-wave components satisfy $g_1-s_1=0\pmod{2}$ and $g_2-s_2=0\pmod{2}$. 
The  normalized out-of-plane spin polarization of each unfolded state
\begin{equation}
P_{z,n\mathbf{K}}(\mathbf{k})
=
\dfrac{W_{n\mathbf{K}}^{\uparrow}(\mathbf{k})-W_{n\mathbf{K}}^{\downarrow}(\mathbf{k})}
{W_{n\mathbf{K}}^{\uparrow}(\mathbf{k})+W_{n\mathbf{K}}^{\downarrow}(\mathbf{k})}.
\end{equation}
Here, $P_{z,n\mathbf K}(\mathbf k)\in[-1,1]$, with positive and negative values corresponding to opposite out-of-plane spin projections.
Within the convention used here, $P_z=\langle\sigma_z\rangle=2\langle S_z\rangle/\hbar$. For visualization, the discrete unfolded states are broadened using a Gaussian function $g_\eta(E-E_{n\mathbf K})$ with broadening $\eta$. The total unfolded spectral function is
\begin{equation}
A_{\mathrm{tot}}(\mathbf{k},E)
=
\sum_{n,\mathbf{K}}
W_{n\mathbf{K}}(\mathbf{k})\,
g_{\eta}\!\left(E-E_{n\mathbf{K}}\right),
\label{eq:spectral_function}
\end{equation}
while the spin-weighted spectral function is defined as
\begin{equation}
A_{\mathrm{spin}}(\mathbf{k},E)
=
\sum_{n,\mathbf{K}}
W_{n\mathbf{K}}(\mathbf{k})
P_{z,n\mathbf{K}}(\mathbf{k})\,
g_{\eta}\!\left(E-E_{n\mathbf{K}}\right).
\label{eq:spin_spectral_function}
\end{equation}
The energy-resolved spin polarization used to color the unfolded band
structure is 
\begin{equation}
P_z(\mathbf{k},E)
=
\frac{A_{\mathrm{spin}}(\mathbf{k},E)}
{A_{\mathrm{tot}}(\mathbf{k},E)},
\label{eq:spectral_spin_polarization}
\end{equation}
The ratio is evaluated only where $A_{tot}(\mathbf{k},E)$ exceeds a small numerical threshold.\\

\begin{figure}[!h]
    \centering
    \includegraphics[width=0.55\linewidth]{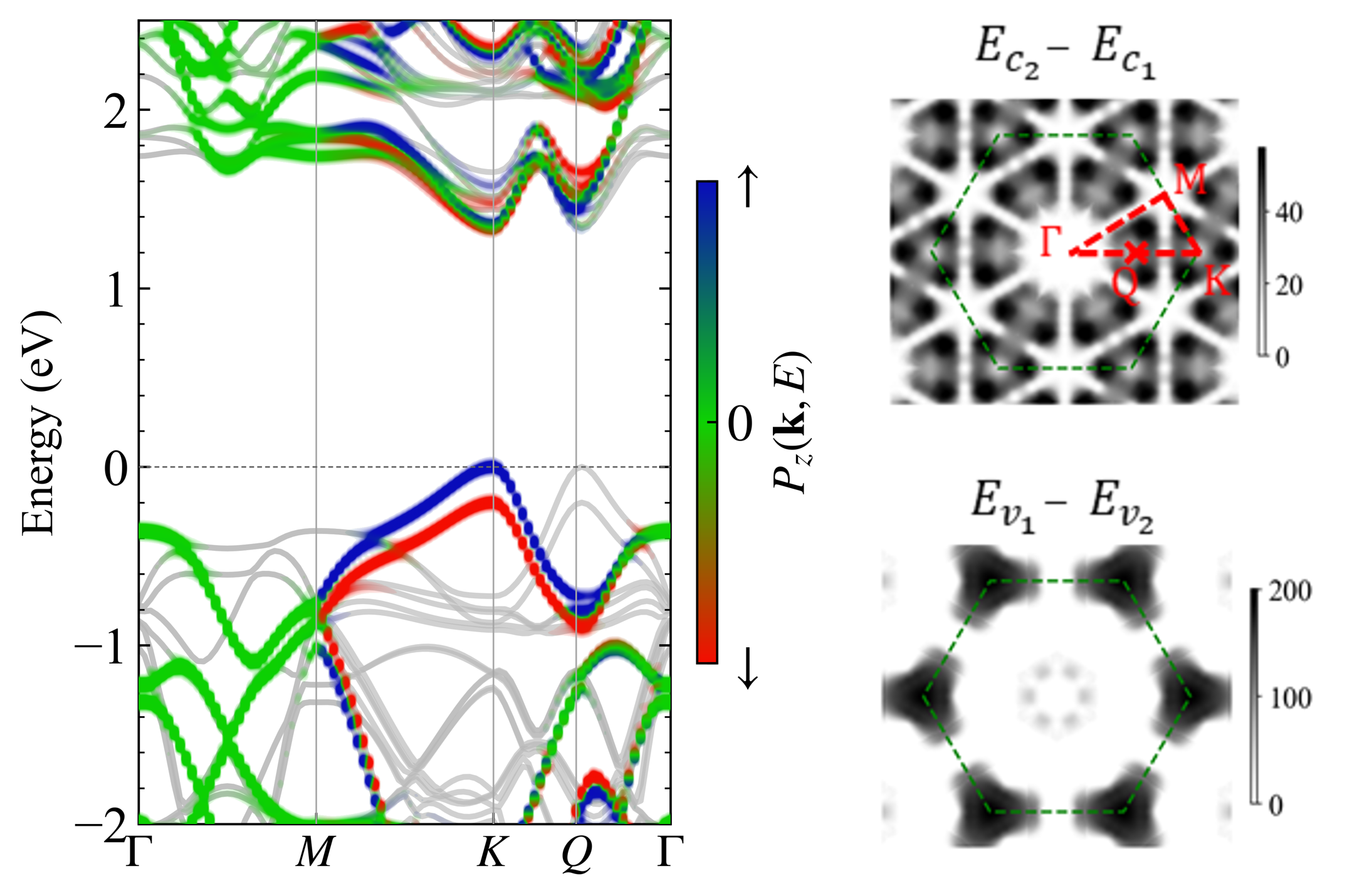}
    \caption{
    Spin-resolved band unfolding and momentum-resolved spin-orbit
    splitting of ML Mo$_{0.75}$W$_{0.25}$Se$_2$ calculated at the DFT+SOC level using a $2\times2\times1$ SC. The left panel shows
    the unfolded electronic structure along the PC $\Gamma$MKQ$\Gamma$ path.  The energy zero is set to the VBM. The color encodes the
    energy-resolved out-of-plane spin polarization $P_z(\mathbf{k},E)$
    defined by Eq.~\eqref{eq:spectral_spin_polarization}, while the spectral intensity is determined by the
    total unfolded spectral function $A_{\mathrm{tot}}(\mathbf{k},E)$ of
    Eq.~\eqref{eq:spectral_function}. Gray curves show the
    underlying folded DFT+SOC bands as a guide to the eye. The right panels show the momentum-resolved conduction-band splitting $\Delta_{\mathrm{SO}}^{c}(\mathbf{k})$ (top) and valence-band splitting $\Delta_{\mathrm{SO}}^{v}(\mathbf{k})$ (bottom) in the PC BZ. The dashed green hexagon marks the first PC BZ.}
    \label{fig:unfolding}
\end{figure}

\noindent
\textbf{Unfolded band structure and spin-orbit splitting}\\
The spin-resolved unfolded band structure of ML Mo$_{0.75}$W$_{0.25}$Se$_2$ is shown in Fig.~\ref{fig:unfolding}.
Unfolding onto the PC BZ allows the band-edge states and their spin character at the $\mathbf{K}^+$ valley to be identified directly.
We denote the two uppermost valence branches by $v_1$ and $v_2$, with energies $E_{v_1}(\mathbf{k})$ and $E_{v_2}(\mathbf{k})$, and the two lowest conduction branches by $c_1$ and $c_2$, with energies $E_{c_1}(\mathbf{k})$ and $E_{c_2}(\mathbf{k})$, respectively.
At the K$^{+}$ valley, $v_1$ corresponds to the valence-band maximum (VBM) and has an out-of-plane spin polarization $P_z=-1$, whereas $c_1$ corresponds to the conduction-band minimum (CBM) and has $P_z=+1$.
The next conduction-band state $c_2$, has $P_z=-1$, \textit{i.e.}, same spin character as $v_1$.
For the momentum-resolved maps shown in the right panels of Fig.~\ref{fig:unfolding}, the  conduction- and valence-band spin-orbit splittings are defined as  $\Delta_{\mathrm{SO}}^{c}(\mathbf{k})=E_{c_2}(\mathbf{k})-E_{c_1}(\mathbf{k})$ and $\Delta_{\mathrm{SO}}^{v}(\mathbf{k})=E_{v_1}(\mathbf{k})-E_{v_2}(\mathbf{k})$.
At the $\mathbf{K}^+$ valley of the PC BZ, the corresponding conduction- and valence-band spin-orbit splittings are $\Delta_{SO}^{c}(K)=18$~meV and $\Delta_{SO}^{v}(K)=200$~meV, respectively.
At the $\mathbf{K}^+$ valley,  the lowest-energy interband transition ($v_1,c_1$) is a spin-forbidden, whereas the higher-energy ($v_1,c_2$) transition is a spin-allowed. 
The $G_0W_0$ quasiparticle calculation and DFT+SOC unfolding provide complementary descriptions of the alloy electronic structure. The resulting band-edge spin ordering is consistent with the assignment of the experimentally observed X$^{\mathrm{D}}$ emission in Mo$_{0.68}$W$_{0.32}$Se$_2$ to a spin-forbidden dark exciton.\\

\clearpage
\newpage

\section{Effective-medium description of magnetic darkening in Mo$_{0.68}$W$_{0.32}$Se$_2$}
\label{sec:SM_discussion}

The theoretical analysis is based on an effective-medium description of the monolayer alloy Mo$_{0.68}$W$_{0.32}$Se$_2$, in which its low-energy electronic structure is assumed to retain the main features of the parent MoSe$_2$ and WSe$_2$ monolayers.
In particular, we assume that well-defined Bloch states can be introduced at the $\mathbf{K}^{\pm}$ valleys, where the relevant excitonic transitions occur.
We further neglect local structural distortions and differences in the spatial extent of the relevant Mo and W orbitals, and describe the alloy by an effective hexagonal lattice with a common lattice constant.
Within this approximation, the electronic states near the $\mathbf{K}^{\pm}$ valleys can be treated using the same symmetry-based framework as in binary S-TMD monolayers~\cite{Xiao2012,Kormanyos2015}.

A $\mathbf{k}\cdot\mathbf{p}$ model for monolayer Mo$_{0.68}$W$_{0.32}$Se$_2$ at the $\mathbf{K}^{\pm}$ points can therefore be constructed analogously to that of the parent S-TMD monolayers.
The simplest four-band model is characterized by: (i) the spin-resolved valence- and conduction-band extrema; (ii) momentum matrix elements between valence- and conduction-band Bloch states of the same spin in the $\mathbf{K}^{\pm}$ valleys, which determine the optical activity of the excitonic transitions; and
(iii) matrix elements coupling these bands to higher conduction and lower valence bands, which provide higher-order corrections within the $\mathbf{k}\cdot\mathbf{p}$ approximation.
We next discuss how these parameters are expected to evolve in Mo$_{0.68}$W$_{0.32}$Se$_2$ compared with those in MoSe$_2$ and WSe$_2$ monolayers.

As a lowest-order approximation, the energy of the $n$th band of the $\mathrm{Mo}_x\mathrm{W}_{1-x}\mathrm{Se}_2$ monolayer at the $\mathbf{K}^{\pm}$ points can be estimated by linear interpolation between the corresponding band energies of the parent compounds:
\begin{align}
\label{eq:weighted} E_n(\mathrm{Mo}_x\mathrm{W}_{1-x}\mathrm{Se}_2) = xE_n(\mathrm{MoSe}_2) + (1-x)E_n(\mathrm{WSe}_2).
\end{align}
Such a linear interpolation provides a simple qualitative description of the evolution of the band-edge energies with composition, although deviations from linearity and band-gap bowing are known to occur in Mo$_x$W$_{1-x}$Se$_2$ alloys~\cite{Tongay2014,Kopaczek2021}.

Within this approximation, the conduction-band spin splitting in the alloy for $x=0.68$ at the $\mathbf{K}^{\pm}$ points becomes substantially smaller in absolute value than the corresponding splittings in pure MoSe$_2$ and WSe$_2$ monolayers.
The reduced conduction-band spin splitting enhances the mixing between bright and dark excitons induced by an in-plane magnetic field.
Consequently, a larger fraction of the optical activity can be transferred from the bright to the dark exciton than in the parent MoSe$_2$ and WSe$_2$ monolayers.
The resulting loss of optical activity of the bright exciton increases with increasing magnetic field.

The momentum matrix elements between valence- and conduction-band Bloch states of the same spin in the $\mathbf{K}^{\pm}$ valleys can likewise be approximated by interpolation between the corresponding values for MoSe$_2$ and WSe$_2$.
These matrix elements determine the optical activity of the bright excitons.
However, within the present model, they do not depend explicitly on the applied in-plane magnetic field and therefore do not themselves account for the magnetic-field-induced darkening.

Finally, we consider higher-order corrections to the four-band $\mathbf{k}\cdot\mathbf{p}$ model.
Changes in the relative positions of the bands modify the contributions of remote bands to the low-energy electronic states and, consequently, their magnetic response~\cite{Bir1974,Kormanyos2015}.
The effective in-plane coupling parameter $g_\parallel$ entering the bright-dark mixing can therefore differ substantially from the corresponding values in the parent MoSe$_2$ and WSe$_2$ monolayers.
A larger $|g_\parallel|$ enhances the magnetic-field-induced mixing of the bright and dark excitons and thus increases the transfer of oscillator strength between the two states.

In addition to this redistribution of optical activity, the bright-like exciton shifts to higher energy with increasing $B_\parallel$.
At finite temperature, this energy shift reduces the thermal population of the bright-like branch and thereby suppresses its photoluminescence intensity.
As shown in the main text, this population effect provides the dominant contribution to the pronounced magnetic darkening observed experimentally.

Interestingly, a large effective in-plane excitonic $g$ factor has recently been reported for a Mo$_{0.7}$W$_{0.3}$Se$_2$ monolayer~\cite{Schindler2026}.
In that work, however, the enhanced photoluminescence $g$ factor was attributed to many-body interactions rather than directly to a single-particle bright-dark mixing parameter.
Nevertheless, the observation highlights the unusually strong in-plane magnetic response of Mo$_x$W$_{1-x}$Se$_2$ alloys close to the conduction-band spin-orbit crossover.

The present model is restricted to neutral bright and dark excitons.
A quantitative description of trions, which are three-particle complexes and involve additional Coulomb and exchange interactions, is beyond the scope of this work.

\newpage
\section{Sensitivity of the extracted $g_\parallel$ factor to the bright-dark exciton splitting}
\label{sec:SM_sensitivity}

\begin{figure}[h!]
    \centering
    \includegraphics[width=0.4\linewidth]{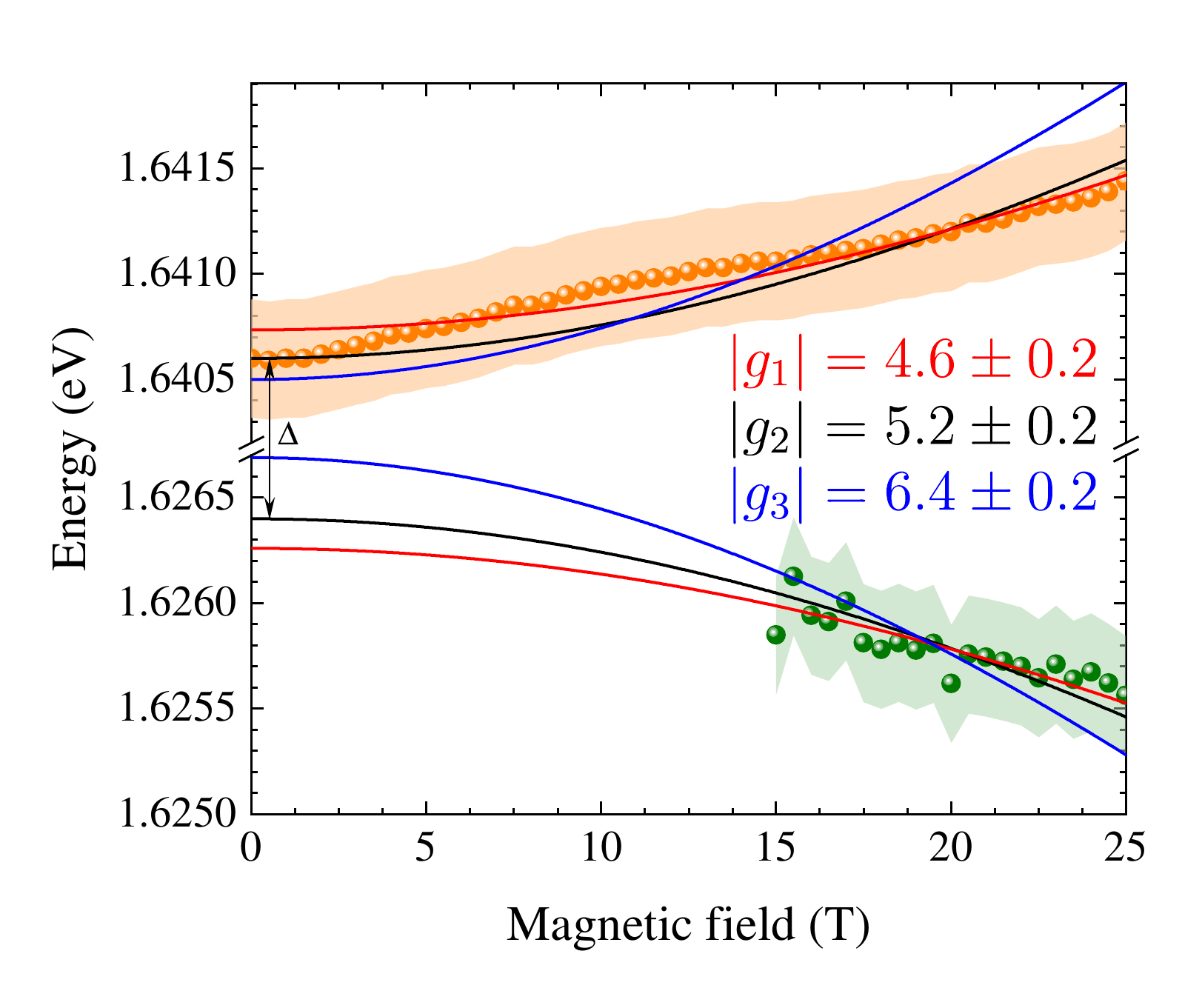}
    \caption{Magnetic-field evolution of the energies of the X$^\mathrm{B}$ and X$^\mathrm{D}$ excitons.
    The shaded areas indicate the uncertainties of the extracted integrated intensities.
    Solid lines show fits based on the bright-dark exciton mixing model described in the main text, obtained using different constraints on the zero-field bright-dark splitting $\Delta$ and the in-plane coupling parameter $|g_\parallel|$.
    }
    \label{fig:g_delta_sensitivity}
\end{figure}

The extracted value of $|g_\parallel|$ is sensitive to the assumed value of the zero-field bright-dark exciton splitting $\Delta$.
To quantify this sensitivity, we performed three fits using different constraints on $\Delta$ and $|g_\parallel|$.
The corresponding fits are shown in Fig.~\ref{fig:g_delta_sensitivity}.
The uncertainty in the extracted transition energies is primarily determined by the experimental energy resolution and spectral sampling.
The energy spacing between adjacent data points in the recorded PL spectra was approximately $0.28~\mathrm{meV}$.

For the main fit, the mean exciton energy $E_0$ was determined by averaging $[E_\mathrm{B}(B_\parallel)+E_\mathrm{D}(B_\parallel)]/2$
over the field range $15$-$25~\mathrm{T}$, where both transitions are experimentally resolved.
Together with the zero-field energy of X$^\mathrm{B}$, this yields $\Delta_1=14.2\pm1.3~\mathrm{meV}$, as described in the main text.
Fixing this value of $\Delta$ in the fit gives $|g_1|=5.2\pm0.2$.

As a second fitting procedure, $\Delta$ was treated as a free parameter, yielding $\Delta_2=14.48\pm0.12~\mathrm{meV}$ and $|g_2|=4.6\pm0.2$.
Conversely, fixing the in-plane coupling parameter to the literature value $|g_3|=6.4$ reported for a Mo$_{0.7}$W$_{0.3}$Se$_2$ monolayer~\cite{Schindler2026}
gives $\Delta_3=13.51\pm0.24~\mathrm{meV}$.

The values of $\Delta_2$ and $\Delta_3$ are consistent with $\Delta_1$ within its uncertainty.
At the same time, the corresponding values of $|g_\parallel|$ differ substantially, demonstrating that the extracted in-plane coupling parameter is highly sensitive to relatively small variations in the assumed zero-field bright-dark exciton splitting.

Overall, the three fitting procedures yield bright-dark splittings in the range of approximately $13.5$-$14.5~\mathrm{meV}$, while the corresponding values of $|g_\parallel|$ span the range $4.6$-$6.4$.
Despite this sensitivity of $|g_\parallel|$, all fits consistently place the zero-field dark-exciton energy at approximately $1.62~\mathrm{eV}$, about $14~\mathrm{meV}$ below the bright exciton.
Thus, the conclusion that the dark exciton forms the lowest-energy excitonic state is robust with respect to the fitting procedure.

\end{document}